\documentclass[pdflatex,sn-mathphys-num,iicol]{sn-jnl}

\usepackage{graphicx}%
\usepackage{multirow}%
\usepackage{amsmath,amssymb,amsfonts}%
\usepackage{amsthm}%
\usepackage{mathrsfs}%
\usepackage[title]{appendix}%
\usepackage{xcolor}%
\usepackage{textcomp}%
\usepackage{manyfoot}%
\usepackage{booktabs}%
\usepackage{algorithm}%
\usepackage{algorithmicx}%
\usepackage{algpseudocode}%
\usepackage{listings}%
\usepackage{subcaption}%
\usepackage{braket}
\usepackage{booktabs}  
\usepackage{siunitx}
\usepackage{longtable}
\newcommand{\nucleus}[2]{^{#1}\textrm{#2}}
\newcommand{\zerotozero}{0^{+}_{1}\to 0^{+}_{1}}
\newcommand{\zerotozerotwo}{0^{+}_{1}\to 0^{+}_{2}}
\newcommand{\zerototwo}{0^{+}_{1}\to 2^{+}_{1}}

\newcommand{\closurekn}{\langle K_{N}\rangle}
\newcommand{\closureln}{\langle L_{N}\rangle}

\newcommand{\abs}[1]{\left|#1\right|}
\newcommand{\psfunit}[1]{\left(10^{#1}\text{y}^{-1}\right)}

\newcommand{\bminusbminus}{\beta^{-}\beta^{-}}
\newcommand{\twonutwobminus}{^{2\nu\bminusbminus}}
\newcommand{\twonutwobplus}{^{2\nu\beta^{+}\beta^{+}}}
\newcommand{\twonuecbplus}{^{2\nu\textrm{EC}\beta^{+}}}
\newcommand{\twonutwoec}{^{2\nu\textrm{ECEC}}}
\newcommand{\zeronutwobminus}{^{0\nu\bminusbminus}}

\newcommand{\zeronuecbplus}{^{0\nu\textrm{EC}\beta^{+}}}
\newcommand{\twobminus}{^{\beta\beta}}

\newcommand{\ecbplus}{^{\textrm{EC}\beta^{+}}}
\newcommand{\twoec}{^{\textrm{ECEC}}}

\newcommand{\eref}[1]{eq.~(\ref{#1})}

\usepackage{longtable}

\theoremstyle{thmstyleone}%
\theoremstyle{thmstyletwo}%

\theoremstyle{thmstylethree}%

\begin{document}

\title[Updated Results for Kinematic Factors in Double Beta Decays]{Updated Results for Kinematic Factors in Double Beta Decays}

\author*[1]{\fnm{S.} \sur{Ghinescu}}\email{stefan.ghinescu@nipne.ro}
\author*[1]{\fnm{S.} \sur{Stoica}}\email{sabin.stoica@cifra-c2unesco.ro}

\affil[1]{\orgname{International Center for Advanced Training and Research in Physics}, \orgaddress{\street{P.O. Box MG-12}, \city{M\u{a}gurele}, \postcode{077125},  \country{Romania}}}

\abstract{Accurate calculations of phase space factors (PSFs), electron energy spectra and angular correlations are essential for designing and interpreting double-beta decay (DBD) experiments. These quantities help maximize sensitivity to potential signals, distinguish between different decay modes and interpret the data. In this work we provide updated results for these kinematic factors for two-neutrino ($2\nu\beta\beta$) and neutrinoless ($0\nu\beta\beta$) decay modes, including electron-emission, positron-emission and electron capture transitions. The calculations are performed with an adapted Dirac-Hartree-Fock-Slater method which allows for orthogonality of the wave functions of electrons and positrons in bound and continuum states and incorporates relevant atomic features such us screening, finite nuclear size, exchange corrections and phase shift effects. We provide tables with updated PSFs calculated both in the closure approximation and using the Taylor expansion method, for a large number of DBD isotopes. We discuss the impact of individual atomic corrections and find that our results are in line with predictions reported in recent literature. In some specific cases we find differences between our PSF values and those previously reported which are worth considering for better prediction and interpretation of DBD data. Then, we provide numerical values for $\nucleus{76}{Ge}$, $\nucleus{100}{Mo}$, $\nucleus{130}{Te}$ and $\nucleus{136}{Xe}$, which are most investigated in current DBD experiments. Similar data for other isotopes are available upon request.}


\maketitle

\section{Introduction}
\label{sec:intro}
Double-beta decay remains an intensively investigated process, the main challenge being the discovery of the hypothetical neutrinoless double-beta decay mode which would be reliable evidence for the existence of beyond Standard Model (BSM) physics. Among other things, its observation would establish lepton number violation (an important condition for leptogenesis), imply that neutrinos are massive Majorana particles and provides insight into the absolute neutrino mass and neutrino mass ordering. Conversely, the absence of a $0\nu\beta\beta$ decay  signal places stringent constraints on a wide range of beyond SM scenarios contributing to this decay mode. There are many reviews on DBD (see for example~\cite{Haxton-PPNP1984,Doi-PTPS1985, Tomoda_RPP_1991, Suhonen_PR_1998, Vogel_ARNPS_2002, Barea_PRC_2009, Vergados_RPP_2012, Dolinski_ARNPS_2019,Blaum-RMP2020,Bossio-JPG2024} and the references therein), where the reader can find comprehensive information about the evolution, present status and perspectives of the field. 

Current DBD experiments have set lower bounds on the $0\nu\bminusbminus$ half-life at the level of $10^{25}$--$10^{26}$~yr~\citep{KamLAND-Zen-PRL2023, EXO-200-PRL2019,GERDA-PRL2020,Adams-COURE-PRL2020} corresponding to the effective neutrino mass in the range of a few tens of meV. Next-generation experiments aim to improve these limits by at least one order of magnitude \citep{abgrall_legend-1000_2021,Adhikari-NEXO-JPG2022,Alfonso-CUPID-JLTP2023,Abdukerim-PANDAXxT-SCPMA2024}, with the goal of fully probing the inverted hierarchy region of the neutrino mass. At this level of sensitivity, experimental efforts require substantial resources, making precise and reliable theoretical inputs essential.

The phase-space factors constitute key ingredients in the double-beta decay investigation for both quantitative predictions and the interpretation of experimental data. Accurate determinations of PSFs, together with related kinematic observables such as single electron and summed-energy electron spectra and angular correlations between emitted electrons or positrons, are increasingly important as experimental precision improves. These quantities enable more reliable decay-rate estimates, facilitate the discrimination between different DBD mechanisms, and are used for detector calibration and data analysis in DBD experiments.

Electron energy spectra provide characteristic signatures for event identification. In the case of $2\nu\beta\beta$, the electron spectral shape is well established, and any deviation from the standard prediction may signal new physics, such as Lorentz invariance violation \cite{Diaz-PRD2014,Nitescu_JPG_2020,Nitescu-PRD2021,Ghinescu-PRD2022}, violation of the Pauli exclusion principle \cite{Barabash-NPB2007}, right-handed neutrino interactions \cite{Deppisch_PRL_2020}, neutrino self-interactions \cite{Deppisch-PRD2020}, existence of sterile neutrinos~\cite{Bolton-PRD2021,Agostini-PLB2021}, or $0\nu\beta\beta$ decay with Majoron emission~\cite{Mohapatra-PLB2000,Kotila_PRC_2015}. For $0\nu\beta\beta$ decay, the summed-energy electron spectrum is expected to appear as a sharp peak at the transition $Q$-value, making precise energy calibration crucial for signal identification. Angular correlations between the emitted leptons provide complementary information, allowing discrimination among decay mechanisms and between signal and background, and play an important role in detector optimization and data analysis.

On the theoretical side the calculation of the PSFs and their associated differential observables has significantly progressed in the last decade. During time, the Fermi functions were calculated with different methods. In the earliest calculation they were built from electron wave functions obtained by taking the square of the ratio between the radial solution of the Schor\"odinger equation for a point charge $Z_f$ and a plane wave evaluated at origin~\cite{Primakoff-RPP1959}. This non-relativistic treatment turned out to be a crude approximation. Later, adopting a relativistic treatment, the Fermi functions were built from radial solutions of a Dirac equation in a Coulomb potential given by a point charge or by a charge uniformly distributed in a finite size nucleus. Using some approximations, the Fermi functions were still obtained in analytical form and they were used in many calculations ~\cite{Doi_PTP_1983,Doi-PTPS1985,Doi_PTP_1992,Tomoda_RPP_1991,Doi_PTP_1993,Suhonen_PR_1998}. As the DBD experiments increase the accuracy in measuring the electron spectra, the theoretical derivation of the Fermi functions improved, as well. The Fermi functions were built with exact electronic wave functions obtained by solving the Dirac equation in a Coulomb potential~\cite{Kotila_PRC_2012} or a realistic Coulomb potential~\cite{Stoica-PRC2013}, including the finite nuclear size correction and screening effects. In this work, we adopt the improved derivation of the Fermi functions used in previous works~\cite{Nitescu-PRC2023,Nitescu-PRC2025}.

We present updated calculations of PSFs, single electron and summed-energy electron spectra as well as angular correlations between the emitted charged leptons, for all nuclei that can undergo double-beta decays. The analysis encompasses the $(2\nu/0\nu)\bminusbminus$, $(2\nu/0\nu)\beta^+\beta^+$, $(2\nu/0\nu)$EC$\beta^+$, and $2\nu$ECEC decay modes, and includes transitions to the ground state (g.s.) as well as to the first excited states $0^+$ and $2^+$. The relevant expressions are derived within the standard closure approximation, assuming the left-handed light-neutrino exchange (LNE) mechanism for $0\nu\beta\beta$ decay. For the $2\nu\bminusbminus$ and $2\nu$ECEC channels, we additionally employ a Taylor expansion method that accounts more accurately for the lepton-energy dependence of DBD rate~\cite{Deppisch_PRL_2020,Nitescu_Universe_2021, Nitescu_Universe_2024,Nitescu_JPG_2024}. The resulting PSFs and differential distributions are expressed in terms of newly introduced nuclear-structure parameters, experimentally accessible, enabling further experimental studies, such as tests of the single-state dominance (SSD) and higher-state dominance (HSD) hypotheses, or alternative determinations of the effective axial-vector coupling constant $g_A^{\mathrm{eff}}$. Both approaches, closure approximation and Taylor expansion method, allow the separation of PSFs from NMEs in the decay rate formulas.

The calculations are performed with an adapted Dirac-Hartree-Fock-Slater (DHFS) method which allows for orthogonality of the wave functions of electrons and positrons in bound and continuum states and incorporates relevant atomic features such us screening, finite nuclear size, exchange corrections and phase shift effects. An improved treatment of the transition energy balance and consequently of the $Q$-values~\cite{Nitescu_JPG_2024} is also implemented.  We discuss the impact of individual atomic corrections and find that our results are in line with recent predictions from literature. Regarding PSFs, in specific cases we find differences between our values and those previously reported, which are worth considering for better prediction and interpretation of DBD data. Numerical electron spectra and angular correlations are also provided for the isotopes $\nucleus{76}{Ge}$, $\nucleus{100}{Mo}$, $\nucleus{130}{Te}$, and $\nucleus{136}{Xe}$ which are most investigated in current DBD experiments, and similar data for other isotopes are available upon request.

The paper is organized as follows. In Section 2, we introduce the theoretical framework adopted in this work, starting with general considerations on the closure approximation and Taylor expansion methods used to derive DBD rates. We then review the expressions for the Q-values of the processes analyzed. Subsections that follow present the formulas of the PSFs, electron spectra and angular correlations for the analyzed decay processes. In Section 3, we describe the computational approach used to obtain the leptonic wave functions and provide tables with PSF values for all nuclei that can undergo double-beta decays. A discussion on the obtained results and comparison with other works from literature is included. Also, we provide supplementary files containing numerical electron spectra and angular correlations for the isotopes $^{76}$Ge, $^{100}$Mo, $^{130}$Te and $^{136}$Xe.

\section{Theoretical Framework}

The calculation of double-beta decay rates requires, in principle, summing over a large number of intermediate nuclear states, a task that becomes computationally demanding, particularly for heavy nuclei. In the closure approximation, the energies of the intermediate states appearing in the energy denominators of the decay rate are replaced by a single average value, referred to as the closure energy $\langle E \rangle$. This approximation significantly simplifies the calculation of nuclear matrix elements (NMEs), as it avoids the explicit treatment of individual excited states and allows the use of the completeness relation for evaluating the contributions of the weak interaction operators. The closure approximation has been shown to be reliable, especially for the $0\nu\beta\beta$ decay mode. However, its accuracy depends on the choice of the closure energy and may introduce systematic uncertainties. Improved accuracy in the determination of the decay-rates is achieved by employing a Taylor expansion method, which explicitly accounts for the dependence of the energy denominators on the energies of the emitted leptons. In this approach, the lepton energies are treated as small parameters relative to a characteristic excitation energy of the intermediate nucleus, and the energy denominators are expanded as a Taylor series up to fourth order in these parameters. This procedure generates additional contributions involving products of different NMEs and their associated PSFs. 

\subsection{Definition of $Q$-value}
\label{subsec:q-values}
$Q$-values represent important input data that significantly influence the calculation of the kinematic quantities. For the ground-state to ground-state $\beta^-\beta^-$ transitions of the nuclei considered in~\cite{Nitescu_JPG_2020}, we use the $Q$-values compiled therein. For all other transitions, the $Q$-values are determined using the calculation formulas given in this section for each decay mode using atomic masses from~\cite{Wang_CPC_2021} and excitation energies from~\cite{ENSDF_2025}. We denote by 
\begin{equation}
    \mathcal{M}(A,Z) = M(A,Z) + Zm_{e}+B(Z)
\end{equation}
the atomic mass of a neutral atom with $Z$ protons and $A-Z$ neutrons. Here $M(A,Z)$ denotes the nuclear mass and the term $B(Z)$ is the binding energy of the electron cloud and it is negative.

The $2\nu\beta^{-}\beta^{-}$ decay can be symbolized as
\begin{equation}
    ^{A}_{Z}X\to^{A}_{Z+2}Y^*+e_1+e_2+\bar{\nu}_1+\bar{\nu}_{2},
\end{equation}
where the $^*$ indicates a possibly excited final nuclear state. The energy balance reads
\begin{align}
\begin{aligned}
    \mathcal{M}(A,Z) &= M(A,Z+2) + E^*\\
    &+Zm_e+B^{*}(Z)\\
    &+\epsilon_{1}+\epsilon_{2}+2m_{e}+\omega_{1}+\omega_{2}
\end{aligned}
\end{align}
where $\epsilon_{1,2}$ and $\omega_{1,2}$ represent the kinetic energies of the electrons and anti-neutrinos, respectively. The final atom is a positive ion of charge $+2e$. The term $E^*$ is the excitation energy of the final nuclear state (equal to 0 in case of g.s. to g.s. transitions). By approximating $B^*(Z)$ with $B(Z+2)$ the sum of the kinetic energies of the four emitted leptons, i.e. the $Q$-value of the process, can be written as
\begin{align}
\label{eq:q_val_twobminus}
\begin{aligned}
    Q^{\bminusbminus} &\equiv \epsilon_{1}+\epsilon_{2}+\omega_{1}+\omega_{2}\\
      &= \mathcal{M}(A,Z)-\mathcal{M}(A,Z+2)-E^*.
\end{aligned}
\end{align}
For $0\nu\beta^{-}\beta^{-}$ in the above formulas one takes $\omega_{1}=\omega_{2}=0$.

The $2\nu\beta^+\beta^+$ decay can be symbolized as
\begin{equation}
    ^A_{Z}X\to^A_{Z+2}Y^*+e^+_1+e^+_2+\nu_1+\nu_2,
\end{equation}
with the same conventions as before. The energy balance reads
\begin{align}
    \begin{aligned}
        \mathcal{M}(A,Z) &= M(A,Z-2)+E^*\\
        &+Zm_e+B^*(Z)\\
        &+\epsilon_{1}+\epsilon_{2}+2m_{e}+\omega_{1}+\omega_{2}.
    \end{aligned}
\end{align}
Here $\epsilon_{1,2}$ and $\omega_{1,2}$ stand for the kinteic energies of the positrons and neutrinos respectively. The final atom is a negative ion of net charge $-2e$ and by approximating $B^*(Z)$ with $B(Z-2)$ we obtain the sum of kinetic energies of the emitted leptons
\begin{align}
\label{eq:q_val_twobplus}
    \begin{aligned}
        Q^{\beta^+\beta^+}&\equiv\epsilon_{1}+\epsilon_{2}+\omega_{1}+\omega_{2}\\
        &=\mathcal{M}(A,Z) - \mathcal{M}(A,Z-2)-4m_{e}-E^*.
    \end{aligned}
\end{align}
When considering the neutrinoless case, formulas remain unchanged and $\omega_1=\omega_{2}=0$.

The EC$\beta^+$ decay can be symbolized as
\begin{equation}
    ^A_ZX\to^A_{Z-2}Y^*+e^++\nu_1+\nu_2.
\end{equation}
In what follows we will consider that the electron is captured from shell $x$. The energy balance reads
\begin{align}
    \begin{aligned}
        \mathcal{M}(A,Z) &= M(A,Z-2)+E^*\\
        &+(Z-1)m_{e}+B_{x}^*(Z-1)\\
        &+\epsilon+m_{e}+\omega_{1}+\omega_{2}
    \end{aligned}
\end{align}
where $\epsilon$ is the kinetic energy of the emitted positron and $\omega_{1,2}$ are the kinetic energies of the neutrinos. The final atom is a negative ion of net charge $-1e$ and its electron cloud is excited, having a hole where the captured electron original resided. The binding energy of the electron cloud is denoted by $B^*_{x}(Z-1)$. By approximating $B(Z-2)-B^{*}_x(Z-1)$ with the binding energy of the captured electron ($-|t_{x}|$) we obtain the $Q$-value of the process
\begin{align}
\label{eq:q_val_ecbplus}
    \begin{aligned}
        Q^{\text{EC}\beta+}-|t_x|&\equiv \epsilon_{1}+\omega_{1}+\omega_{2}\\
        &= \mathcal{M}(A,Z)-\mathcal{M}(A,Z-2)\\
        &-2m_e-E^*
    \end{aligned}
\end{align}
The neutrinoless mode yields similar formulas with $\omega_{1}=\omega_{2}=0$ (hence the positron takes the full available energy).

Finally, the $2\nu$ECEC transition with captures from shells $x$ and $y$ can be symbolized as
\begin{equation}
    ^{A}_{Z}X\to^{A}_{Z-2}Y^*+\nu_{1}+\nu_{2}
\end{equation}
The energy balance reads
\begin{align}
    \begin{aligned}
        \mathcal{M}(A,Z)&=M(A,Z-2)+E^*\\
        &+(Z-2)m_e+B_{x,y}(Z-2)\\
        &+\omega_{1}+\omega_{2}
    \end{aligned}
\end{align}
Here $\omega_{1,2}$ are the energies of the neutrinos and the final atom is neutral. The electron cloud is excited, having holes in the shells $x$ and $y$. By approximating $B_{x,y}(Z-2)-B(Z-2)$ with $|t_x|+|t_y|$ (see \citep{Nitescu_JPG_2024} for a rigorous derivation), the definition of the $Q$-value becomes
\begin{align}
\label{eq:q_val_twoec}
    \begin{aligned}
        Q^{\text{ECEC}}-|t_x|-|t_y|&\equiv\omega_1+\omega_{2}\\
        &=\mathcal{M}(A,Z)-\mathcal{M}(A,Z-2)\\
        &-E^*
    \end{aligned}
\end{align}

\subsection{$2\nu\bminusbminus$}
\label{subsec:twonutwobeta}
In the case of DBD with double electron emission, the differential decay rate with respect to the angle between the emitted electrons is given by~\cite{Doi_PTP_1983, Haxton-PPNP1984,Doi-PTPS1985, Tomoda_RPP_1991, Doi_PTP_1993}
\begin{equation}
    \label{eq:Gamma_2charged}
    \frac{d\Gamma\twonutwobminus}{d(\cos\theta)} = \frac{\Gamma\twonutwobminus}{2}\left(1+K\twonutwobminus\cos\theta \right)
\end{equation}
where~\cite{Doi_PTP_1983, Haxton-PPNP1984,Doi-PTPS1985, Tomoda_RPP_1991, Doi_PTP_1993}
\begin{equation}
    \label{eq:K_2charged}
    K\twonutwobminus = \frac{\Lambda\twonutwobminus}{\Gamma\twonutwobminus}
\end{equation}
is the angular correlation coefficient. 

\paragraph{Closure Approximation}
In the closure approximation the terms $\Gamma\twonutwobminus$ and $\Lambda\twonutwobminus$ can be factorized as follows\cite{Doi_PTP_1983, Haxton-PPNP1984,Doi-PTPS1985, Tomoda_RPP_1991, Doi_PTP_1993}:
\begin{align}
\label{eq:gam_lam_closure_2nubb}
    \begin{aligned}
        \begin{Bmatrix}
            \Gamma\twonutwobminus\\
            \Lambda\twonutwobminus
        \end{Bmatrix}
        =
        g_A^{4}\left|M\twonutwobminus\right|^2 \ln(2)
        \begin{Bmatrix}
            G\twonutwobminus\\
            H\twonutwobminus
        \end{Bmatrix}
    \end{aligned}
\end{align}
where $g_A$ is the vector-axial constant, $M\twonutwobminus$ is the nuclear matrix element and $G\twonutwobminus$ and $H\twonutwobminus$ are phase space factors, their forms depending on the state of the final nucleus.

Combining the results of \cite{Doi_PTP_1983, Haxton-PPNP1984,Doi-PTPS1985, Tomoda_RPP_1991, Doi_PTP_1993} with the recent work \cite{Nitescu-PRC2025}, the PSFs for the $0^+_{1}\to0^+_1$ transition appearing in~\eref{eq:gam_lam_closure_2nubb} and including the radiative and exchange corrections are defined as 
\begin{align}
\scriptscriptstyle
    \begin{aligned}
        &\begin{Bmatrix}
            G\twonutwobminus\\
            H\twonutwobminus
        \end{Bmatrix} =\frac{\tilde{A}^{2}\left(G_{F}|V_{ud}|\right)^{4}}{96\pi^{7}\ln(2)} \\
        &\times\int_{0}^{Q\bminusbminus}d\epsilon_1 \int_{0}^{Q\bminusbminus-\epsilon_{1}}d\epsilon_2\int_{0}^{Q\bminusbminus-\epsilon_{1}-\epsilon_{2}} d\omega_{1} \\
        &\times R(\epsilon_{1},Q^{\bminusbminus})R(\epsilon_{2},Q^{\bminusbminus}) (\epsilon_{1}+m_{e})\\
        &\times p_{1}(\epsilon_{2}+m_e)p_{2} \omega_{1}^{2}\omega_2^{2} \begin{Bmatrix}
            f_{11}^{0}\\
            f_{11}^{1}
        \end{Bmatrix}\\
        &\times \begin{Bmatrix}
            \left(\closurekn^{2}+\closureln^{2}+\closurekn \closureln\right)\\
            \frac{2}{3}\left(\closurekn^{2}+\closureln^{2}+\frac{5}{2}\closurekn \closureln\right)
        \end{Bmatrix}
    \end{aligned}
\end{align}

In the expressions above, $\epsilon_{1,2}$ and $p_{1,2}$ are the kinetic energies and momenta of the emitted electrons, $\omega_{1}$ and $\omega_2 = Q\twobminus-\epsilon_{1}-\epsilon_{2}-\omega_{1}$ are the energies of the antineutrinos. The value of $Q^{\bminusbminus}$ is given by \eref{eq:q_val_twobminus}. The factors $\closurekn$ and $\closureln$ are given by~\cite{Doi_PTP_1983, Haxton-PPNP1984,Doi-PTPS1985, Tomoda_RPP_1991, Doi_PTP_1993}
\begin{align}
\label{eq:kn_ln_closure}
    \begin{aligned}
        \closurekn = &\frac{1}{\epsilon_{1} +m_e +\omega_{1} + \langle E_{n}\rangle -E_{I}}+\\
        &\frac{1}{\epsilon_{2} +m_e +\omega_{2} + \langle E_{n}\rangle -E_{I}}, \\
        \closureln = &\frac{1}{\epsilon_{2} +m_e +\omega_{1} + \langle E_{n}\rangle -E_{I}}+\\
        &\frac{1}{\epsilon_{1} +m_e +\omega_{2} + \langle E_{n}\rangle -E_{I}}.
    \end{aligned}
\end{align}
We denote by $E_{n}$ the energies of $1^{+}_{n}$ states of the intermediary nucleus and by $\langle E_{n}\rangle$ the average excitation energy, while $E_I$ denotes the energy of the initial nucleus. Furthermore, $f_{11}^{0}$ and $f_{11}^{1}$ are defined by~\cite{Tomoda_RPP_1991}
\begin{align}
    \begin{aligned}
        f_{11}^{0} &= \left|f^{-1-1}\right|^{2}+\left|f_{11}\right|^{2} \\
        &+ \left|{f^{-1}}_{1}\right|^{2} + \left|{f_1}^{-1}\right|^{2},\\
        f_{11}^{1} &= -2\Re\left(f^{-1-1}f^{*}_{11} + {f^{-1}}_{1}{f_1}^{-1*}\right)
    \end{aligned}
\end{align}
where
\begin{align}
    \label{eq:fermi_func_def}
    \begin{aligned}
        f^{-1-1} &= g_{-1}(\epsilon_{1})g_{-1}(\epsilon_{2}),\\
        f_{11} &= f_{1}(\epsilon_{1})f_{1}(\epsilon_{2}),\\
        {f^{-1}}_{-1} &= g_{-1}(\epsilon_{1})f_{1}(\epsilon_{2}),\\
        {f_{1}}^{-1} &= f_{1}(\epsilon_{1})g_{-1}(\epsilon_{2}),\\
    \end{aligned}
\end{align}
in terms of the large $g_{-1}(\epsilon)\equiv g_{-1}(\epsilon, R)$ and small $f_{1}(\epsilon)\equiv f_{1}(\epsilon_{1}, R)$ components of the continuum electron wave functions, which account for the exchange correction (see next section). We employ the usual approximation of the giant Gamow-Teller resonance $\tilde{A}=\frac{1}{2}W_{0}+\langle E_{n}\rangle -E_{I}\approx 1.12A^{1/2}~$MeV.

The radiative correction is accounted for through the $R$ term, which is given by~\cite{Hayen-RMP2018, Nitescu-PRC2025}
\begin{align}
    \begin{aligned}
        R(\epsilon,\epsilon_{\mathrm{max}}) = 1 + \frac{\alpha}{2\pi}g(\epsilon,\epsilon_{\mathrm{max}})
    \end{aligned}
\end{align}
where $\alpha$ is the fine structure constant and $g(\epsilon, \epsilon_{\mathrm{max}})$ is given by~\cite{Hayen-RMP2018, Nitescu-PRC2025}
\begin{align}
    \begin{aligned}
        &g(\epsilon, \epsilon_{\mathrm{max}}) = 3\ln(m_p)-\frac{3}{4}-\frac{4}{\beta}\mathrm{Li}_2\left(\frac{2\beta}{1+\beta}\right) \\
        &+ \frac{\tanh^{-1}\beta}{\beta}\\
        &\times\left[2(1+\beta^2)+\frac{\epsilon_{\mathrm{max}}-\epsilon}{6(\epsilon+m_e)^2}-4\tanh^{-1}\beta\right] \\
        &+ 4\left(\frac{\tanh^{-1}\beta}{\beta}-1\right)\\
        &\times\left[\frac{\epsilon_{\mathrm{max}}-\epsilon}{3(\epsilon+m_e)}-\frac{3}{2}+\ln\left[2\left(\epsilon_{\mathrm{max}}-\epsilon\right)\right]\right]
    \end{aligned}
\end{align}

The same formulas are valid for the $\zerotozerotwo$ transition. The only modification is the total kinetic energy release as discussed in the previous section. 

The single electron-energy spectra, summed-energy electron spectra and the angular correlation between the emitted electrons/positrons are derived from the PSF expressions, as follows~\cite{Doi_PTP_1983, Haxton-PPNP1984,Doi-PTPS1985, Tomoda_RPP_1991, Doi_PTP_1993}
\begin{align}
    \begin{aligned}
        &\begin{Bmatrix}
            \frac{dG\twonutwobminus}{d\epsilon_1}\\
           \frac{dH\twonutwobminus}{d\epsilon_1}
        \end{Bmatrix}
        = \frac{\tilde{A}^{2}\left(G_{F}|V_{ud}|\right)^{4}}{96\pi^{7}\ln(2)} \\ &\times R(\epsilon_1, Q^{\bminusbminus})(\epsilon_1+m_e)p_1\\
        &\times\int_{0}^{Q^{\bminusbminus}-\epsilon_1}d\epsilon_2\int_{0}^{Q^{\bminusbminus}-\epsilon_{1}-\epsilon_2}d\omega_{1}\\
        &\times\omega_{1}^{2}\omega_2^{2} \begin{Bmatrix}
            f_{11}^{0}\\
            f_{11}^{1}
        \end{Bmatrix}\\
        &\times \begin{Bmatrix}
            \left(\closurekn^{2}+\closureln^{2}+\closurekn \closureln\right)\\
            \frac{2}{3}\left(\closurekn^{2}+\closureln^{2}+\frac{5}{2}\closurekn \closureln\right)
        \end{Bmatrix}
    \end{aligned}
\end{align}
\begin{align}
    \begin{aligned}
        &\frac{dG\twonutwobminus}{dT} = \frac{\tilde{A}^{2}\left(G_{F}|V_{ud}|\right)^{4}}{96\pi^{7}\ln(2)} \frac{T}{Q^{\bminusbminus}}\\
        &\times \int_{0}^{Q^{\bminusbminus}}dV\int_{0}^{Q^{\bminusbminus}-T}d\omega_{1} \\
        &\times R(\epsilon_1,Q^{\bminusbminus}) R(\epsilon_{2},Q^{\bminusbminus})\\
        & \times (\epsilon_1+m_e)p_1(\epsilon_2+m_e)p_2 \\
        &\times f_{11}^{0}\omega_{1}^{2}\omega_2^{2}\\
        &\times \left(\closurekn^{2}+\closureln^{2}+\closurekn \closureln\right).
    \end{aligned}
\end{align}
In the last equation, we used the same change of variables as in \cite{Simkovic-PRC2018} to define the summed electron energy spectrum, namely 
\begin{align}
    \begin{aligned}
        \begin{Bmatrix}
            T\\
            V
        \end{Bmatrix}&=
        \begin{Bmatrix}
            \epsilon_1+\epsilon_2\\
            Q\frac{\epsilon_2}{\epsilon_1+\epsilon_2}
        \end{Bmatrix}\\
        \begin{Bmatrix}
            \epsilon_1\\
            \epsilon_2
        \end{Bmatrix}
        &=
        \begin{Bmatrix}
            T\left(1-\frac{V}{Q^{\bminusbminus}}\right)\\
            \frac{TV}{Q^{\bminusbminus}}
        \end{Bmatrix}
    \end{aligned}
\end{align}
The electron angular correlation is defined as~\cite{Doi_PTP_1983, Haxton-PPNP1984,Doi-PTPS1985, Tomoda_RPP_1991, Doi_PTP_1993}
\begin{align}
    \begin{aligned}
        \alpha\twonutwobminus = \frac{dH\twonutwobminus/d\epsilon_1}{dG\twonutwobminus/d\epsilon_1}.
    \end{aligned}
\end{align}

The PSF integrals for the $\zerototwo$ transition are given by~\cite{Doi_PTP_1983, Haxton-PPNP1984,Doi-PTPS1985, Tomoda_RPP_1991, Doi_PTP_1993} (again accounting for the radiative and exchange corrections as above)
\begin{align}
    \begin{aligned}
        &\begin{Bmatrix}
            G\twonutwobminus\\
            H\twonutwobminus
        \end{Bmatrix} =\frac{\tilde{A}^{6}\left(G_{F}|V_{ud}| \right)^{4}}{96\pi^{7}\ln(2)} \\
        &\times \int_{0}^{Q^{\bminusbminus}}d\epsilon_1 \int_{0}^{Q^{\bminusbminus}-\epsilon_{1}}d\epsilon_2\int_{0}^{Q^{\bminusbminus}-\epsilon_{1}-\epsilon_{2}} d\omega_{1} \\
        &\times R(\epsilon_{1},Q^{\bminusbminus})R(\epsilon_{2},Q^{\bminusbminus}) \\
        &\times(\epsilon_{1}+m_{e})p_{1}(\epsilon_{2}+m_e)p_{2} \\
        &\times \omega_{1}^{2}\omega_2^{2} \left(\closurekn^2-\closureln^2\right) \begin{Bmatrix}
            f_{11}^{0}\\
            \frac{1}{3}f_{11}^{1}
        \end{Bmatrix}
    \end{aligned}
\end{align}

\paragraph{Taylor Expansion Method}
Within the Taylor Expansion method, the terms $\Gamma\twonutwobminus$ and $\Lambda\twonutwobminus$ for the $\zerotozero$ transition are given by~\citep{Deppisch_PRL_2020,Nitescu_Universe_2021, Nitescu_Universe_2024}
The terms appearing in its definition are given by
\begin{align}
    \begin{aligned}
        \Gamma\twonutwobminus &= \Gamma_{0} + \Gamma_{2} + \Gamma_{22} + \Gamma_{4},\\
        \Lambda\twonutwobminus &= \Lambda_{0} + \Lambda_{2} + \Lambda_{22} + \Lambda_{4},
    \end{aligned}
\end{align}
where
\begin{align}
\label{eq:gamma_lambda_factorization_taylor}
    \begin{aligned}
        \Gamma_{N} &= g_{A}^{4}\mathcal{M}_{N}\twonutwobminus G_{N}\twonutwobminus,\\
        \Lambda_{N} &= g_{A}^{4}\mathcal{N}_{N}\twonutwobminus H_{N}\twonutwobminus,
    \end{aligned}
\end{align}
with $N=0,2,22,4$. The phase space factors are given by~~\citep{Deppisch_PRL_2020,Nitescu_Universe_2021, Nitescu_Universe_2024}
\begin{align}
\label{eq:g_h_def_taylor}
    \begin{aligned}
        &\begin{Bmatrix}
            G_N\twonutwobminus\\
            H_N\twonutwobminus
        \end{Bmatrix}
        =\frac{(G_{F}\left|V_{ud}\right|^{4})}{8\pi^{7}}\\
        &\times \int_{0}^{Q^{\bminusbminus}}d\epsilon_{1}\int_{0}^{Q^{\bminusbminus}-\epsilon_{1}}d\epsilon_{2}\int_{0}^{Q^{\bminusbminus}-\epsilon_{1}-\epsilon_{2}}d\omega_{1}\\
        &\times R(\epsilon_1,Q^{\bminusbminus})R(\epsilon_1,Q^{\bminusbminus})\\
        &\times(\epsilon_{1}+m_{e})p_{1}(\epsilon_{2}+m_{e})p_{2}\omega_{1}^{2}\omega_{2}^{2}\mathcal{A}_{N}
        \begin{Bmatrix}
            f_{11}^0\\
            f_{11}^1
        \end{Bmatrix}.
    \end{aligned}
\end{align}

The factors $\mathcal{A}_{N}$ are given by~~\citep{Deppisch_PRL_2020,Nitescu_Universe_2021, Nitescu_Universe_2024}
\begin{align}
\label{eq:a_factors_taylor}
\begin{aligned}
    \mathcal{A}_{0} &= 1,\hspace{1cm} \mathcal{A}_{2} = \frac{\epsilon_{K}^{2}+\epsilon_{L}^{2}}{(2m_{e})^{2}} \\
    \mathcal{A}_{22} &= \frac{\epsilon_{K}^2\epsilon_{L}^2}{(2m_e)^4} \hspace{1cm} \mathcal{A}_{4} = \frac{\epsilon_{K}^{4}+\epsilon_{L}^{4}}{(2m_{e})^{4}},
\end{aligned}
\end{align} 
and the quantities $\mathcal{M}$ and $\mathcal{N}$ depend on the nuclear matrix elements and can be found in e.g. \cite{Nitescu_Universe_2021}.

The factors $\epsilon_{K}$ and $\epsilon_{L}$ depend on the energies of the emitted lepton as follows~\citep{Deppisch_PRL_2020,Nitescu_Universe_2021, Nitescu_Universe_2024}
\begin{align}
\begin{aligned}
    \epsilon_{K} &= \frac{1}{2}(\epsilon_{2}+\omega_{2}-\epsilon_{1}-\omega_{1}) \\
    \epsilon_{L} &= \frac{1}{2}(\epsilon_{1}+\omega_{2}-\epsilon_{2}-\omega_{1}). 
\end{aligned}
\end{align}

In the case of the $\zerototwo$ transition, the differential decay rate can be written as in \eref{eq:Gamma_2charged}, but with the following modification~\citep{NEMO3_EPJC_2025}
\begin{align}
    \begin{aligned}
        \Gamma\twonutwobminus &=  \Gamma_{22} + \Gamma_{6},\\
        \Lambda\twonutwobminus &= \Lambda_{22} + \Lambda_{6}.
    \end{aligned}
\end{align}
Here, eqs.~(\ref{eq:gamma_lambda_factorization_taylor}), and~(\ref{eq:g_h_def_taylor}) hold formally and the factors $\mathcal{A}_{N}$ are given by~\cite{NEMO3_EPJC_2025}
\begin{align}
\begin{aligned}
    \mathcal{A}_{22} &= \frac{\left(\epsilon_{K}^2-\epsilon_{L}^{2}\right)^{2}}{(2m_e)^4},\\
    \mathcal{A}_{6} &= 2\frac{\left(\epsilon_{K}^2-\epsilon_{L}^{2}\right)^{2}\left(\epsilon_{K}^2+\epsilon_{L}^{2}\right)}{(2m_e)^6}.
\end{aligned}
\end{align}

\subsection{$0\nu\bminusbminus$}
The differential decay rate for the $\zerotozero$ transition in the LNE mechanism, produced by left-handed currents, can be written as~\cite{Doi-PTPS1985, Doi_PTP_1993, Tomoda_RPP_1991}
\begin{align}
\begin{aligned}
    \frac{d\Gamma\zeronutwobminus}{d(\cos\theta)} &= g_A^4 \left|M\zeronutwobminus\right|^2 \frac{\left|m_{\beta\beta}\right|^2}{m_e^{2}}\\
    &\times\frac{G\zeronutwobminus}{2}\left(1+K\zeronutwobminus\cos\theta\right)
\end{aligned}
\end{align}
where~\cite{Doi-PTPS1985, Doi_PTP_1993, Tomoda_RPP_1991}
\begin{equation}
    K\zeronutwobminus = \frac{H\zeronutwobminus}{G\zeronutwobminus}
\end{equation}
and~\cite{Doi-PTPS1985, Doi_PTP_1993, Tomoda_RPP_1991}
\begin{align}
\begin{aligned}
&\begin{Bmatrix}
    G\zeronutwobminus \\
    H\zeronutwobminus
\end{Bmatrix}
= \frac{\left(G_{F}|V_{ud}|\right)^{4}}{32\pi^{5}R^2}\int_{0}^{Q^{\bminusbminus}}d\epsilon_1\\
&\times(\epsilon_{1}, \epsilon_{2})(\epsilon_{1}+m_{e})p_{1}(\epsilon_{2}+m_e)p_{2}
        \begin{Bmatrix}
            f_{11}^{0}\\
            f_{11}^{1}
        \end{Bmatrix}
\end{aligned}
\end{align}
The effective Majorana neutrino mass is given by~\cite{Doi-PTPS1985, Doi_PTP_1993, Tomoda_RPP_1991}
\begin{equation}
    m_{\beta\beta}=\sum_{k=1}^{3}U_{ek}^{2}m_{k},
\end{equation}
where $U_{ek}$ $(k=1,2,3)$ are the elements of the Pontecorvo-Maki-Nakagawa-Sakata matrix and $m_{k}$ are the values of the neutrino masses.

\subsection{$2\nu\beta^+\beta^+$ and $0\nu\beta^+\beta^+$}
The formulas for PSFs, electron spectra and angular correlation in the closure approximation presented above for double electron emission channels remain valid in the case of the $2\nu/0\nu\beta^+\beta^+$ transitions~\cite{Doi_PTP_1983,Doi-PTPS1985, Tomoda_RPP_1991, Doi_PTP_1993}. However, the expressions of the $Q$-values are taken from~\eref{eq:q_val_twobplus}. From computational point of view, the wavefunctions entering the definitions of the Fermi functions in \eref{eq:fermi_func_def} are obtained from the solution of the Dirac equation for a positron in the daughter atom field. Moreover, the positron wavefunctions are not corrected for the exchange effect. We also mention that the Taylor expansion method has not (yet) been applied to the double positron emission transitions. since the experimental investigations of these transitions are not (yet) developed.

\subsection{$2\nu$EC$\beta^{+}$}
Within the closure approximation, the decay rate is given by~\cite{Doi_PTP_1992}
\begin{equation}
\label{eq:Gamma_2nuecbetaplus}
    \Gamma\twonuecbplus = g_{A}^4 \left|M\twonuecbplus\right|^{2} G\twonuecbplus
\end{equation}
where $M\twonuecbplus$ is NME and, for $\zerotozero$ and $\zerotozerotwo$ transitions~\cite{Doi_PTP_1992},
\begin{align}
    \begin{aligned}
        &G\twonuecbplus = \frac{2\tilde{A}^{2}\left(G_{F}\left|V_{ud}\right|\right)^{4}}{48\pi^{5}}m_e \\
        &\times\sum_{x}\mathcal{B}_{x}^{2}\int_{0}^{Q\ecbplus-\left|t_{x}\right|}d\epsilon_{p}\\
        &\times\int_{0}^{Q\ecbplus-\abs{t_x}-\epsilon_{p}}d\omega_{1} \\
        &\times \left[g_{-1}^{2}(\epsilon_{p}) + f_{1}^{2}(\epsilon_{p})\right] \\
        &\times\left(\closurekn^{2}+\closureln^{2}+\closurekn\closureln\right)\\
        &\times \omega_{1}^{2}\omega_{2}^{2}p_{p}(\epsilon_{p}+m_{e}),
    \end{aligned}
\end{align}
where~\cite{Doi_PTP_1992},
\begin{equation}
    \mathcal{B}_{x}^{2} = \frac{1}{4\pi m_{e}^{3}}\left[g_{x}^{2}(R) + f_{x}^{2}(R)\right]
\end{equation}
is the probability of an electron from shell $x$ to be localized within the nucleus. Here, $g_{x}$ and $f_{x}$ are the large and small components of the captured electron radial wave function and $R=1.2A^{1/3}$. The sum over shells goes over all occupied major shells, but we consider only $s$ subshells, meaning $\kappa=-1$.  The total energy of the captured electron is given by $e_{x}=m_e-\abs{t_{x}}$, where $t_{x}<0$ is the binding energy. Similarly to the notation in the previous sections, $p_{p}$ and $\epsilon_{p}$ denote the momentum and kinetic energy of the emitted positron. Finally, $\closurekn$ and $\closureln$ have the same formal definition as in \eref{eq:kn_ln_closure} with $\epsilon_{1}\to \epsilon_{p}$ and $\epsilon_{2}+m_e\to-e_{x}$.  For this channel, the definition of $Q\ecbplus$ is given by~\eref{eq:q_val_ecbplus}.

\subsection{$0\nu$EC$\beta^+$}
The decay rate for this channel is given by~\cite{Doi_PTP_1993}
\begin{equation}
    \Gamma\zeronuecbplus = g_A^4 \left|M\zeronuecbplus\right|^2 \frac{\left|m_{\beta\beta}\right|^2}{m_e^{2}}G\zeronuecbplus
\end{equation}
where~\cite{Doi_PTP_1993}
\begin{align}
\begin{aligned}
    &G\zeronuecbplus = \frac{(G_F|V_{ud}|)^4}{8\pi^{3}R^2}m_e^5  \\
    &\times \sum_{x} B_{x}^2[g_{-1}(\epsilon_p,R)^2+f_1(\epsilon_p,R)^2]\\
    &\times p_p(\epsilon_p+m_e)
\end{aligned}
\end{align}

\subsection{$2\nu$ECEC mode}
\label{subsec:2nu2ec}
We limit the discussion to $\zerotozero$ transitions for this mode. 

\paragraph{Closure Approximation}
The decay rate formula in the closure approximation can be also factorized in the same way as for the previous decay transitions~\cite{Doi_PTP_1992}:
\begin{equation}
    \Gamma\twonutwoec = g_A^{4}\left|M\twonutwoec\right|^{2} \ln(2) G\twonutwoec,
\end{equation}
where $M\twonutwoec$ is the nuclear matrix element and $G\twonutwoec$ is the phase space factor, given by~\cite{Doi_PTP_1993}
\begin{align}
\label{eq:G_2nuECEC}
    \begin{aligned}
        &G\twonutwoec = \frac{2\tilde{A}^{2}\left(G_F\abs{V_{ud}}^{2}\right)}{48\pi^{3}\ln2} m_e^4 \\
        &\times\sum_{x,y} \mathcal{B}_x^{2} \mathcal{B}_y^2  \int_{0}^{Q\twoec-\abs{t_x}-\abs{t_y}}d\omega_{1} \\
        & \times \left[\closurekn^{2}+\closureln^{2}+\closurekn\closureln\right]\omega_{1}^{2}\omega_{2}^{2},
        \end{aligned}
\end{align}
for the $\zerotozero$ and $\zerotozerotwo$ transitions. Here $Q\twoec$ is given by \eref{eq:q_val_twoec} and in the formulae of $\closurekn$ and $\closureln$ the following replacements should be performed: $\epsilon_{1}+m_e\to -(m_e-\abs{t_x})$ and $\epsilon_{2}+m_e\to -(m_e-\abs{t_y})$. Again we consider only $\kappa=-1$ sub-shells.  

\paragraph{Taylor Expansion Method}
The Taylor expansion method has been proposed for this mode in~\cite{Nitescu_JPG_2024}. Also in this case, the decay rate includes also the additional terms
\begin{equation}
    \Gamma\twonutwoec = g_A^{4}(\Gamma_{0}+\Gamma_2+\Gamma_{22}+\Gamma_4)
\end{equation}
where~\cite{Nitescu_JPG_2024}
\begin{equation}
    \Gamma_N = g_A^{4}\mathcal{M}_N\twonutwoec G_N\twonutwoec, N=0,2,22,4.
\end{equation}
The phase space factor $G_N\twonutwoec$ are defined as~\cite{Nitescu_JPG_2024}
\begin{equation}
    G_{N}\twonutwoec = \frac{(G_{F}|V_{ud}|)^4}{2\pi^{3}}\sum_{x,y}\mathcal{B}_{x}^{2}\mathcal{B}_{y}^{2}I_{N,xy}
\end{equation}
where 
\begin{equation}
    I_{N,xy} = \int_{0}^{Q\twoec-|t_x|-|t_y|}d\omega_{1} \omega_{1}^{2}\omega_{2}^2\mathcal{A}_{N}
\end{equation}
where $\mathcal{A}_{N}$ are defined by \eref{eq:a_factors_taylor} with the replacement $\epsilon_{1,2}\to -(m_e-|t_{x,y}|)$. All other expressions are identical to the ones in the previous sections.


\section{Results and Discussion}
In calculating PSFs the main ingredients are the Fermi functions which encode the interaction between the emitted electrons/positrons and the electrostatic field of the final atom. They can be expressed in terms of electron and positron radial wave functions evaluated on the nuclear surface, which we obtain with the DHFS self-consistent method and using the \verb|RADIAL| computing package \citep{Salvat-CPC2019}. Our method allows an accurate treatment of relevant atomic and nuclear effects such as scattering, exchange, radiative, finite nuclear size and phase shift effects. We particularly mention the exchange corrections which were introduced in a more consistent way in refs.~\cite{Nitescu-PRC2023,Nitescu_AIP_2024,NEMO3_EPJC_2025}, but for particular cases. By modifying the last iteration of the DHFS method, we also ensured orthogonality between the electron wave functions of bound and continuous states, which results in a better description of the electron spectra especially in the low-energy region of electron spectra. 

A particular treatment is used for  $\beta^+\beta^+$ and EC$\beta^+$ decays where the final atom is a negative ion, which prevents the direct use of the DHFS method to model its electron cloud\footnote{The potentials for the positron emission decays (eqs.~(\ref{eq:potential_z2},\ref{eq:potential_z1}) are proposed by O.V. Ni\c{t}escu, but are unpublished as of the writing of this manuscript}. To avoid this issue, we construct the potential felt by the scattering positron(s) as follows. Suppose that the final atom has $Z$ protons in its nucleus and $Z+2$ electrons in its cloud (in the case of $\beta^+\beta^+$) decay. We approximate the potential acting on the scattering positrons due to the electron cloud to be
\begin{align}
    \label{eq:potential_z2}
	\begin{aligned}
		V_{\text{el}}(Z;Z+2) &= V_{\text{el}}(Z;Z)\\
		&+\left(V_{\text{el}}(Z;Z)-V_{\text{el}}(Z;Z-2)\right),
	\end{aligned}
\end{align}
where $V_{\text{el}}(Z;Z)$ is the electrostatic potential generated by the electron cloud of the neutral atom with atomic number $Z$ and $V_{\text{el}}(Z;Z-2)$ is the electrostatic potential of the positive ion with $Z$ protons in its nucleus and $Z-2$ electrons in its cloud. For the EC$\beta^+$, the potential felt by the scattering positron due to the electron cloud of the final atom (with $Z$ protons and $Z+1$ electrons) is 
\begin{align}
    \label{eq:potential_z1}
	\begin{aligned}
		V_{\text{el}}(Z;Z+1) &= V_{\text{el}}(Z;Z)\\
		&+\left(V_{\text{el}}(Z;Z)-V_{\text{el}}(Z;Z-1)\right).
	\end{aligned}
\end{align}

The PSFs for $2\nu\bminusbminus$ and $0\nu\bminusbminus$ decays are shown in  Tables~\ref{tab:2nu_2bminus_0_to_0},~\ref{tab:2nu_2bminus_0_to_2} and, respectively,~\ref{tab:0nu_2bminus_0_to_0}. The PSFs for $2\nu\beta^+\beta^+$ and $0\nu\beta^+\beta^+$ are shown in Tables~\ref{tab:2nu_2bplus_0_to_0} and~\ref{tab:0nu_2bplus_0_to_0}, respectively. The $2\nu$EC$\beta^+$ and $0\nu$EC$\beta^+$ PSFs are shown in Tables~\ref{tab:2nu_ECbplus_0_to_0} and~\ref{tab:0nu_ECbplus_0_to_0}, respectively. Finally, the $2\nu$ECEC PSFs are shown in Table~\ref{tab:2nu_2EC_0_to_0}. Along with the PSFs we provide, in all tables, for each nucleus and transition the respective $Q$-value. We considered all nuclei for which the $\zerotozero$ transition is energetically allowed.

Our calculations, extended to a large number of isotopes, confirm the general trend of the atomic effects reported in the previous calculations performed for particular isotopes. For $\bminusbminus$ decay mode, we confirm earlier findings~\cite{Nitescu-PRC2025} that the atomic exchange effects are significantly enhanced relative to single $\beta$ decay. This enhancement arises from the fact that the exchange corrections for the two emitted electrons are multiplicative. Consequently, the single-electron energy spectrum exhibits a pronounced increase at low electron energies. 

Radiative corrections increase the total decay rate by about 5\% but have a negligible effect on spectral shapes. When combined with atomic exchange effects these corrections act constructively, shifting the peak of the summed-energy electron spectra to lower energies.

The role of electron phase shifts in calculating angular correlations aligns with prior analyses~\cite{Nitescu_Universe_2024}. For all studied nuclei, the angular correlation coefficient increases in both the $2\nu\bminusbminus$ and $0\nu\bminusbminus$ decay modes for $\zerotozero$ transitions. For the $2\nu\bminusbminus$ our values of the angular correlation coefficient are larger than previously reported ones (e.g.~\cite{Kotila_PRC_2012,Stoica-PRC2013}) by 2--40\% for $\zerotozero$ transitions and 10--80\% for transitions to excited states. In the neutrinoless case, the increases are in the range 1--12\% for $\zerotozero$ transitions while for transitions to excited state our values are smaller by 1--40\% than previously reported ones. 

In $2\nu\beta^+\beta^+$ decay, improved atomic cloud modeling of the DHFS method produces effects analogous to those in $2\nu\bminusbminus$ decay, yielding small deviations from the conventional Thomas-Fermi screening used in calculations with other methods. Radiative corrections here reflect those observed in $2\nu\bminusbminus$ decay. For $2\nu$ECEC decay, the impact of updated PSFs varies with nuclear mass. For light nuclei the deviations from simplified atomic screening (limited to $K$- and $L_1$- shell capture) are negligible, reflecting a balance between reduced decay rates (from accurate screening) and enhanced rates (from higher-shell capture). For medium/heavy nuclei the cancellation is incomplete, with higher shell capture dominating. This increases the decay rate linearly with atomic number $Z$, reaching $\sim$ 10\% for the heaviest nuclei. In EC$\beta^+$ decay mode, the combined effects on PSFs reflect a combination of $\beta^+\beta^+$ and $2\nu$ECEC behaviors. Improved atomic modeling slightly reduces the total decay-rate by lowering capture probability, while positron emission remains largely unaffected. Radiative corrections increase the total decay-rate, but to a lesser extent than in $\beta^+\beta^+$) decay. Including captures from all $\kappa=-1$ orbitals further increases the rate, though less markedly than in $2\nu$ECEC.


We first analyze the results for the $2\nu\bminusbminus$ transitions computed within the closure approximation. Our PSFs for the $\zerotozero$ transitions are generally $4-10$\% larger than the previously reported values~\cite{Kotila_PRC_2012,Stoica-PRC2013}. A similar discrepancy is noted in~\cite{Nitescu-PRC2025}, and most of this effect appears to arise from the radiative correction. An outlier is $\nucleus{154}{Sm}$, where our PSF is about $40$\% larger than the value in~\cite{Kotila_PRC_2012}, likely because of a substantial difference in the $Q$-value. 

The values of $H\twonutwobminus$ PSFs for the $\zerotozero$ transitions generally agree with earlier results to within about 10\%. While the radiative corrections increases the absolute values, this effect is offset by including the phase shift. 

Similar conclusions apply to transitions to excited $0^{+}$ states. Our PSFs for $\zerotozerotwo$ transitions are $10-40$\% larger than those in recent work~\cite{Stoica_FP_2019}. This increase is again due to improved screening, the inclusion of exchange and radiative corrections, and updated decay energies. 

Regarding the results obtained with the Taylor expansion method for the $2\nu\bminusbminus$ PSFs, our findings agree within 2\% with those reported in recent publications~\cite{Nitescu-PRC2025} and~\cite{NEMO3_EPJC_2025}. These observations indicate that the DHFS approach to modeling the atomic cloud produces wave functions for emitted electrons comparable to those obtained using the Thomas-Fermi screening method employed in earlier studies. The inclusion of radiative corrections increases the PSF values by a few percent. Additionally, the exchange correction has a relatively minor effect, typically within a few percent, depending on transition and becomes more significant at lower energies~\cite{Nitescu-PRC2025}.

The same discussion applies to $0\nu\bminusbminus$ transitions. However, because of the smaller overall phase space factor, involving a single integral, the differences between our PSFs and previously reported values are reduced with respect to the differences obtained in the $2\nu\bminusbminus$ case. For $\zerotozero$ transitions, we obtain $G\zeronutwobminus$ and $H\zeronutwobminus$ larger by $0.4-9$\%. For $\zerototwo$ transitions, our values agree well with earlier results only for decay energies $Q^{\bminusbminus}>1$ MeV. Below this energy, the differences exceed 10\%, in some cases reaching a factor of 2. We attribute this trend to both the treatment of the atomic electron cloud and to the exchange corrections.

For both modes of the $\bminusbminus$ decay we have made available data files containing electron energy spectra and angular correlations for $\nucleus{76}{Ge}$, $\nucleus{100}{Mo}$, $\nucleus{130}{Te}$ and $\nucleus{136}{Xe}$ which can be found at~\cite{DoubleBetaDecay_Data_2025}.The structure of the files is thoroughly explained in the \verb|README| file in the repository.

For $2\nu\beta^+\beta^+$ decays, our PSFs are generally $5-16$\% higher than those reported in~\cite{Stoica-PRC2013}, and by $13-46\%$ higher than those in~\cite{Kotila_PRC_2013}. This is primarily due to the updated $Q$-values used in our calculations and the improved treatment of the atomic electron cloud. In contrast, for the neutrinoless decays, the discrepancies are much smaller, our PSFs are typically only a few percent higher than previously reported values.

In the cases of EC$\beta^+$ and ECEC decays, captures from all $s$-wave shells were considered, as in~\cite{Nitescu_Universe_2024} and~\cite{Nitescu_JPG_2024}. Furthermode, the DHFS approach for modeling the atomic electron cloud provides significantly more accurate electron binding energies than those obtained with Thomas-Fermi screening. Consequently, our PSFs for $2\nu/0\nu$EC$\beta^+$ decays differ considerably from the previously published ones, with no clear systematic pattern. On the other hand, the $G\twonutwoec$ values presented here are very similar to those from~\cite{Nitescu_Universe_2024} for the closure approximation. Finally the PSFs obtained using the Taylor expansion method for $2\nu$ECEC in $\nucleus{124}{Xe}$ differ by only a few percent from recent results reported in~\cite{Nitescu_JPG_2024}.

\onecolumn
\begin{longtable}{
l
S[round-precision=3,table-format=1.3]
S[round-precision=2,table-format=5.2e1]
S[round-precision=2,table-format=5.2e1]
S[round-precision=2,table-format=5.2e1]
S[round-precision=2,table-format=5.2e1]
S[round-precision=2,table-format=5.2e1]
}
\caption{
\label{tab:2nu_2bminus_0_to_0}Values of $2\nu\bminusbminus$ PSFs in the closure and Taylor approximations for $0^{+}_{1}\to0^{+}_{1}$ and $0^{+}_{1}\to0^{+}_{2}$ transitions.}\\
\toprule
Nucleus & \ensuremath{$Q$\text{-value}}  & \ensuremath{G^{2\nu\bminusbminus}\psfunit{-21}}  & \ensuremath{G_{0}\psfunit{-21}}  & \ensuremath{G_{2}\psfunit{-21}}  & \ensuremath{G_{22}\psfunit{-21}}  & \ensuremath{G_{4}\psfunit{-21}} \\ \midrule
\multirow{2}{*}{$^{46}$Ca} & 0.9886589999999996  & 0.04875697506253016  & 0.04869137817651544  & 0.0024906073455300396  & 3.922E-05  & 1.582E-04 \\ 
 & -1.6223410000000005  &    &    &    &    &   \\ \hline
\multirow{2}{*}{$^{48}$Ca} & 4.26807  & 16179.775584181853  & 15929.155406327445  & 13636.338670838055  & 3289.209899838026  & 14808.200626640908 \\
 & 1.2707599999999997  & 0.37635634488058334  & 0.37546946614747373  & 0.0313697775409904  & 7.875E-04  & 0.003255228678559291 \\ \hline
\multirow{2}{*}{$^{70}$Zn} & 0.997287999999986  & 0.12354831993300736  & 0.1234225914826814  & 0.0065076910275217665  & 1.071E-04  & 4.234E-04 \\
 & -0.2183330000000141  &    &    &    &    &   \\ \hline
\multirow{2}{*}{$^{76}$Ge} & 2.039059  & 51.207147258093414  & 51.0712356532312  & 10.865054254575625  & 0.6797029353483366  & 2.8801011897914304 \\
 & 0.9167799999999999  & 0.07584582856169549  & 0.07592897589928668  & 0.003404340205860712  & 4.804E-05  & 1.873E-04 \\ \hline
\multirow{2}{*}{$^{80}$Se} & 0.13396800000000803  & 7.020E-08  & 7.045E-08  & 6.874E-11  & 2.425E-14  & 8.181E-14 \\
 & -1.186541999999992  &    &    &    &    &   \\ \hline
\multirow{2}{*}{$^{82}$Se} & 2.9979  & 1698.2502583170635  & 1692.4851865417947  & 754.7697771984223  & 96.92057887588595  & 424.11242560048805 \\
 & 1.5102  & 4.899394639640688  & 4.898574095061213  & 0.5840775733163517  & 0.02099928665309409  & 0.08632400006809067 \\ \hline
\multirow{2}{*}{$^{86}$Kr} & 1.257423840000003  & 1.3118424979122405  & 1.3112330138177677  & 0.10947822463680444  & 0.0028207580783014543  & 0.01129163498355127 \\
 & -0.8485761599999968  &    &    &    &    &   \\ \hline
\multirow{2}{*}{$^{94}$Zr} & 1.144747000000003  & 0.8819534626186237  & 0.881509531160357  & 0.06147031495586457  & 0.001333192705429872  & 0.005273371441133723 \\
 & -0.5969029999999971  &    &    &    &    &   \\ \hline
\multirow{2}{*}{$^{96}$Zr} & 3.356028999999995  & 7326.699241960865  & 7286.202130029464  & 4066.279848022639  & 649.0603695541514  & 2849.4162806327945 \\
 & 2.207898999999995  & 191.28821350042273  & 190.75638943639524  & 47.74490528486369  & 3.505415774400945  & 14.862158322686389 \\ \hline
\multirow{2}{*}{$^{98}$Mo} & 0.10889100000000326  & 3.650E-08  & 3.650E-08  & 2.344E-11  & 5.543E-15  & 1.848E-14 \\
 & -1.2132689999999968  &    &    &    &    &   \\ \hline
\multirow{2}{*}{$^{100}$Mo} & 3.0344  & 3508.444762179009  & 3494.4574781929405  & 1610.1355413302842  & 213.66643410114443  & 931.3521942327795 \\
 & 1.904095  & 64.88828374206797  & 64.8367869315557  & 12.205653150961105  & 0.6839933531881618  & 2.849279258392424 \\ \hline
\multirow{2}{*}{$^{104}$Ru} & 1.2993600000000005  & 3.4248340234627563  & 3.4248819285041576  & 0.3068979119530829  & 0.008496409779211057  & 0.03382141193585262 \\
 & -0.034229999999999317  &    &    &    &    &   \\ \hline
\multirow{2}{*}{$^{110}$Pd} & 2.01785  & 147.08656468306063  & 147.20292945258552  & 31.054283332156867  & 1.9460450981274247  & 8.111671763536238 \\
 & 0.5447800000000003  & 0.00525974967621208  & 0.005249123068830597  & 8.409E-05  & 4.543E-07  & 1.651E-06 \\ \hline
\multirow{2}{*}{$^{114}$Cd} & 0.5448010000000068  & 0.006374839749156928  & 0.006371057669740366  & 1.023E-04  & 5.532E-07  & 2.011E-06 \\
 & -1.4084649999999934  &    &    &    &    &   \\ \hline
\multirow{2}{*}{$^{116}$Cd} & 2.8135  & 2937.6062730472877  & 2931.0356581783626  & 1175.4776957720846  & 136.9858667288656  & 588.7316018807974 \\
 & 1.056636  & 0.9611951678292923  & 0.961389471199487  & 0.0571196351104091  & 0.001085395060598427  & 0.004201080004116843 \\ \hline
\multirow{2}{*}{$^{122}$Sn} & 0.37332900000001246  & 4.881E-04  & 4.872E-04  & 3.687E-06  & 9.651E-09  & 3.388E-08 \\
 & -0.9840719999999876  &    &    &    &    &   \\ \hline
\multirow{2}{*}{$^{124}$Sn} & 2.292667000000001  & 605.0228199975264  & 603.0529260033227  & 163.28364268144875  & 13.052178171933843  & 55.15678375107789 \\
 & 0.6353840000000013  & 0.024146156060155164  & 0.024089488178217825  & 5.231E-04  & 3.807E-06  & 1.404E-05 \\ \hline
\multirow{2}{*}{$^{128}$Te} & 0.8667402700000093  & 0.3033229277442138  & 0.30257318286377116  & 0.012199202674058506  & 1.600E-04  & 6.046E-04 \\
 & -0.7162357299999909  &    &    &    &    &   \\ \hline
\multirow{2}{*}{$^{130}$Te} & 2.52751  & 1637.2889199949184  & 1634.187128512625  & 535.1382001685796  & 51.332785273220345  & 217.8790579378003 \\
 & 0.7339899999999999  & 0.08597054648816756  & 0.08586072344321988  & 0.002488935986380652  & 2.384E-05  & 8.868E-05 \\ \hline
\multirow{2}{*}{$^{134}$Xe} & 0.8241675699999905  & 0.2492303833135278  & 0.24917702277532047  & 0.0091014046981354  & 1.086E-04  & 4.077E-04 \\
 & -0.9363874300000096  &    &    &    &    &   \\ \hline
\multirow{2}{*}{$^{136}$Xe} & 2.45813  & 1536.7279213097047  & 1531.907132967523  & 475.63250566218187  & 43.45816272273818  & 183.96624077371666 \\
 & 0.8791610000000001  & 0.4068969341544568  & 0.4066066005801368  & 0.01688019026300317  & 2.273E-04  & 8.571E-04 \\ \hline
\multirow{2}{*}{$^{142}$Ce} & 1.4171589999999996  & 24.458845366447534  & 24.42412456398068  & 2.597784414750391  & 0.08615073877014323  & 0.34206022456942514 \\
 & -0.8003250000000002  &    &    &    &    &   \\ \hline
\multirow{2}{*}{$^{146}$Nd} & 0.07044400000000314  & 1.047E-08  & 1.048E-08  & 2.820E-12  & 2.798E-16  & 9.239E-16 \\
 & -2.1405559999999966  &    &    &    &    &   \\ \hline
\multirow{2}{*}{$^{148}$Nd} & 1.9280009999999892  & 350.40736989485333  & 350.5339701443916  & 68.2343254535326  & 4.005742970749026  & 16.41274073288024 \\
 & 0.5035409999999891  & 0.011523765111336392  & 0.0115039243037968  & 1.580E-04  & 7.398E-07  & 2.641E-06 \\ \hline
\multirow{2}{*}{$^{150}$Nd} & 3.3713849999999947  & 38907.41643250928  & 38762.60132513989  & 22114.604987098228  & 3664.887449994313  & 15861.026204093614 \\
 & 2.6309209999999945  & 4640.661882832716  & 4629.737190976526  & 1644.1137334184316  & 172.25964910104895  & 726.6977050210223 \\ \hline
\multirow{2}{*}{$^{154}$Sm} & 1.2507939999999944  & 13.362437479893863  & 13.365681611459438  & 1.1138995309229345  & 0.02935278772592192  & 0.1140832969945224 \\
 & 0.5701266999999944  & 0.034963199819080505  & 0.0350138630711018  & 6.125E-04  & 3.647E-06  & 1.315E-05 \\ \hline
\multirow{2}{*}{$^{160}$Gd} & 1.7303650000000053  & 212.22189631549855  & 212.09664566699553  & 33.424654834986605  & 1.6153887648569976  & 6.522688248780335 \\
 & 0.45042300000000535  & 0.007644932257119596  & 0.007650026600509977  & 8.364E-05  & 3.168E-07  & 1.123E-06 \\ \hline
\multirow{2}{*}{$^{170}$Er} & 0.6564149999999936  & 0.18261857604561124  & 0.18266191490537337  & 0.004222821606570269  & 3.300E-05  & 1.202E-04 \\
 & -0.41293500000000616  &    &    &    &    &   \\ \hline
\multirow{2}{*}{$^{176}$Yb} & 1.0851059999999997  & 9.865354884098705  & 9.853841774712377  & 0.6200698217498133  & 0.012591337269066  & 0.04778813256750519 \\
 & -0.06483400000000028  &    &    &    &    &   \\ \hline
\multirow{2}{*}{$^{186}$W} & 0.4914300000000003  & 0.04117146005172871  & 0.041254384725003104  & 5.356E-04  & 2.411E-06  & 8.554E-06 \\
 & -0.5695699999999997  &    &    &    &    &   \\ \hline
\multirow{2}{*}{$^{192}$Os} & 0.4062220000000016  & 0.012893585087060499  & 0.012906967100513683  & 1.151E-04  & 3.577E-07  & 1.255E-06 \\
 & -0.7889469999999985  &    &    &    &    &   \\ \hline
\multirow{2}{*}{$^{198}$Pt} & 1.0502969999999987  & 17.852010032150563  & 17.813414185500505  & 1.050589720735841  & 0.02021409188740674  & 0.07595370936373178 \\
 & -0.3512230000000012  &    &    &    &    &   \\ \hline
\multirow{2}{*}{$^{204}$Hg} & 0.4196669999999976  & 0.02543127202119283  & 0.025397744769790304  & 2.414E-04  & 8.031E-07  & 2.817E-06 \\
 & -1.1630330000000024  &    &    &    &    &   \\ \hline
\multirow{2}{*}{$^{232}$Th} & 0.8372649999999994  & 12.557835646940825  & 12.539295759184341  & 0.4716114575821397  & 0.005978993750226929  & 0.021788551985293458 \\
 & 0.14584499999999945  & 4.612E-05  & 4.616E-05  & 5.300E-08  & 2.231E-11  & 7.456E-11 \\ \hline
\multirow{2}{*}{$^{238}$U} & 1.1445840000000025  & 165.00720628941752  & 164.97644209172637  & 11.517226849467825  & 0.2652045465021488  & 0.9906101166057437 \\
 & 0.20311400000000246  & 6.013E-04  & 5.997E-04  & 1.342E-06  & 1.085E-09  & 3.663E-09 \\ \hline
\midrule
Nucleus & \ensuremath{$Q$\text{-value}}  & \ensuremath{H^{2\nu\bminusbminus}\psfunit{-21}}  & \ensuremath{H_{0}\psfunit{-21}}  & \ensuremath{H_{2}\psfunit{-21}}  & \ensuremath{H_{22}\psfunit{-21}}  & \ensuremath{H_{4}\psfunit{-21}} \\ \midrule
\multirow{2}{*}{$^{46}$Ca} & 0.9886589999999996  & -0.015288137138745357  & -0.015277242837199553  & -6.876E-04  & -9.103E-06  & -3.898E-05 \\
 & -1.6223410000000005  &    &    &    &    &   \\ \hline
\multirow{2}{*}{$^{48}$Ca} & 4.26807  & -12175.059758902908  & -12006.855348237277  & -9514.678542230415  & -2128.729305169161  & -9617.482795266 \\
 & 1.2707599999999997  & -0.14448304481186083  & -0.144264080360338  & -0.010655538423349025  & -2.294E-04  & -9.929E-04 \\ \hline
\multirow{2}{*}{$^{70}$Zn} & 0.997287999999986  & -0.03700304076780956  & -0.0369915972334783  & -0.001706408714976391  & -2.332E-05  & -9.882E-05 \\
 & -0.2183330000000141  &    &    &    &    &   \\ \hline
\multirow{2}{*}{$^{76}$Ge} & 2.039059  & -26.168459382953795  & -26.108291005658096  & -4.948557842797196  & -0.27070390891172913  & -1.1832299440209892 \\
 & 0.9167799999999999  & -0.02087740249181312  & -0.02086800286999504  & -8.147E-04  & -9.473E-06  & -3.990E-05 \\ \hline
\multirow{2}{*}{$^{80}$Se} & 0.13396800000000803  & -2.106E-09  & -2.103E-09  & -1.635E-12  & -4.089E-16  & -1.570E-15 \\
 & -1.186541999999992  &    &    &    &    &   \\ \hline
\multirow{2}{*}{$^{82}$Se} & 2.9979  & -1075.8031320825469  & -1071.4045759687856  & -432.2384930125282  & -49.81874248993922  & -221.04576581568742 \\
 & 1.5102  & -2.019487970352605  & -2.016541646747215  & -0.21177425676251999  & -0.006502886427202927  & -0.02798144401277235 \\ \hline
\multirow{2}{*}{$^{86}$Kr} & 1.257423840000003  & -0.4652552777684561  & -0.46499928266971824  & -0.03403206194318798  & -7.336E-04  & -0.0031255680376299617 \\
 & -0.8485761599999968  &    &    &    &    &   \\ \hline
\multirow{2}{*}{$^{94}$Zr} & 1.144747000000003  & -0.28455982703876553  & -0.28442899577456204  & -0.017260194270313448  & -3.119E-04  & -0.0013172568435034513 \\
 & -0.5969029999999971  &    &    &    &    &   \\ \hline
\multirow{2}{*}{$^{96}$Zr} & 3.356028999999995  & -4836.46890990532  & -4815.8343242378205  & -2433.031784892397  & -350.3521904916293  & -1556.3469213728442 \\
 & 2.207898999999995  & -100.5293013445329  & -100.36208988872441  & -22.3306992121085  & -1.4372413528342431  & -6.2709419581060795 \\ \hline
\multirow{2}{*}{$^{98}$Mo} & 0.10889100000000326  & -6.858E-10  & -6.869E-10  & -3.227E-13  & -4.972E-17  & -1.807E-16 \\
 & -1.2132689999999968  &    &    &    &    &   \\ \hline
\multirow{2}{*}{$^{100}$Mo} & 3.0344  & -2198.9034380578423  & -2191.198488743918  & -910.8190769905483  & -108.26135705837365  & -478.4705151620043 \\
 & 1.904095  & -30.84655287375395  & -30.817224619838644  & -5.128814506081354  & -0.24838074780716965  & -1.0751171042200482 \\ \hline
\multirow{2}{*}{$^{104}$Ru} & 1.2993600000000005  & -1.2145881007750643  & -1.213268841028675  & -0.09480152464158641  & -0.0021967153741398035  & -0.009296631497094027 \\
 & -0.034229999999999317  &    &    &    &    &   \\ \hline
\multirow{2}{*}{$^{110}$Pd} & 2.01785  & -72.2535909932781  & -72.15530138880025  & -13.479837363902968  & -0.7332614222763686  & -3.174278374126424 \\
 & 0.5447800000000003  & -8.087E-04  & -8.087E-04  & -1.103E-05  & -4.649E-08  & -1.882E-07 \\ \hline
\multirow{2}{*}{$^{114}$Cd} & 0.5448010000000068  & -9.689E-04  & -9.680E-04  & -1.317E-05  & -5.553E-08  & -2.240E-07 \\
 & -1.4084649999999934  &    &    &    &    &   \\ \hline
\multirow{2}{*}{$^{116}$Cd} & 2.8135  & -1751.8084408487014  & -1747.633063660367  & -628.250466559369  & -65.03798200731826  & -285.1816578666024 \\
 & 1.056636  & -0.2816051063456797  & -0.2815416135037973  & -0.014548047482741812  & -2.260E-04  & -9.420E-04 \\ \hline
\multirow{2}{*}{$^{122}$Sn} & 0.37332900000001246  & -4.746E-05  & -4.744E-05  & -2.974E-07  & -5.847E-10  & -2.313E-09 \\
 & -0.9840719999999876  &    &    &    &    &   \\ \hline
\multirow{2}{*}{$^{124}$Sn} & 2.292667000000001  & -317.96345894442675  & -317.5763864672085  & -76.32797517115996  & -5.330288288001738  & -23.125444816897687 \\
 & 0.6353840000000013  & -0.004266525481003569  & -0.004264643442550031  & -7.912E-05  & -4.512E-07  & -1.827E-06 \\ \hline
\multirow{2}{*}{$^{128}$Te} & 0.8667402700000093  & -0.07267627727701749  & -0.07261810613724648  & -0.002517704114261625  & -2.646E-05  & -1.090E-04 \\
 & -0.7162357299999909  &    &    &    &    &   \\ \hline
\multirow{2}{*}{$^{130}$Te} & 2.52751  & -910.4835860962693  & -908.7828492357977  & -264.76290592928643  & -22.35322307100683  & -97.33931195619344 \\
 & 0.7339899999999999  & -0.01747489259307005  & -0.017485207120968835  & -4.332E-04  & -3.282E-06  & -1.344E-05 \\ \hline
\multirow{2}{*}{$^{134}$Xe} & 0.8241675699999905  & -0.05659416336790236  & -0.05651917424662319  & -0.0017681624288511239  & -1.687E-05  & -6.924E-05 \\
 & -0.9363874300000096  &    &    &    &    &   \\ \hline
\multirow{2}{*}{$^{136}$Xe} & 2.45813  & -836.4038095841468  & -834.9973955420876  & -230.56128912586712  & -18.495897439779284  & -80.22340244405144 \\
 & 0.8791610000000001  & -0.09837375131241531  & -0.0983242659432482  & -0.003505524891927609  & -3.783E-05  & -1.560E-04 \\ \hline
\multirow{2}{*}{$^{142}$Ce} & 1.4171589999999996  & -8.90684626381534  & -8.899475687395439  & -0.8249717383273723  & -0.02276088929942755  & -0.09573645132006049 \\
 & -0.8003250000000002  &    &    &    &    &   \\ \hline
\multirow{2}{*}{$^{146}$Nd} & 0.07044400000000314  & -3.429E-11  & -3.455E-11  & 5.765E-16  & 1.839E-19  & 2.750E-18 \\
 & -2.1405559999999966  &    &    &    &    &   \\ \hline
\multirow{2}{*}{$^{148}$Nd} & 1.9280009999999892  & -160.83614763893266  & -160.71684014431276  & -27.470336146840104  & -1.3828848287271935  & -5.89469364879558 \\
 & 0.5035409999999891  & -0.0015019434943837901  & -0.0015000100262203399  & -1.716E-05  & -6.095E-08  & -2.434E-07 \\ \hline
\multirow{2}{*}{$^{150}$Nd} & 3.3713849999999947  & -24906.240832505242  & -24832.533846278704  & -12766.3455473541  & -1895.608906099138  & -8293.573351658235 \\
 & 2.6309209999999945  & -2601.366398334302  & -2596.3145740694304  & -820.1810695697475  & -75.35702504716419  & -326.6413145875947 \\ \hline
\multirow{2}{*}{$^{154}$Sm} & 1.2507939999999944  & -4.341211005011204  & -4.337663074245119  & -0.31267143537371467  & -0.0067769317376547185  & -0.02823466153304429 \\
 & 0.5701266999999944  & -0.005198756320380649  & -0.005199705156744518  & -7.638E-05  & -3.468E-07  & -1.392E-06 \\ \hline
\multirow{2}{*}{$^{160}$Gd} & 1.7303650000000053  & -89.08034237915489  & -88.99118370885326  & -12.26098479530017  & -0.5015383935684903  & -2.1215978467295664 \\
 & 0.45042300000000535  & -8.533E-04  & -8.527E-04  & -7.669E-06  & -2.171E-08  & -8.582E-08 \\ \hline
\multirow{2}{*}{$^{170}$Er} & 0.6564149999999936  & -0.03084612360558094  & -0.03081226539386565  & -5.974E-04  & -3.597E-06  & -1.442E-05 \\
 & -0.41293500000000616  &    &    &    &    &   \\ \hline
\multirow{2}{*}{$^{176}$Yb} & 1.0851059999999997  & -2.7519926540707322  & -2.7496879400501064  & -0.14818596025644484  & -0.0024282312781449037  & -0.009961090886839349 \\
 & -0.06483400000000028  &    &    &    &    &   \\ \hline
\multirow{2}{*}{$^{186}$W} & 0.4914300000000003  & -0.004807826236870421  & -0.004810386497602042  & -5.067E-05  & -1.692E-07  & -6.638E-07 \\
 & -0.5695699999999997  &    &    &    &    &   \\ \hline
\multirow{2}{*}{$^{192}$Os} & 0.4062220000000016  & -0.0011646785275727764  & -0.0011577039985700712  & -8.180E-06  & -1.825E-08  & -7.087E-08 \\
 & -0.7889469999999985  &    &    &    &    &   \\ \hline
\multirow{2}{*}{$^{198}$Pt} & 1.0502969999999987  & -4.689515923705282  & -4.689620197179597  & -0.23481667543457244  & -0.0036087396971349692  & -0.014668019401810471 \\
 & -0.3512230000000012  &    &    &    &    &   \\ \hline
\multirow{2}{*}{$^{204}$Hg} & 0.4196669999999976  & -0.0023202918451069626  & -0.0023147136078553966  & -1.732E-05  & -4.094E-08  & -1.590E-07 \\
 & -1.1630330000000024  &    &    &    &    &   \\ \hline
\multirow{2}{*}{$^{232}$Th} & 0.8372649999999994  & -2.4961641953826046  & -2.501118367918895  & -0.07787621862025963  & -7.602E-04  & -0.0030115973910511975 \\
 & 0.14584499999999945  & -4.435E-07  & -4.469E-07  & -1.011E-10  & 7.489E-15  & 3.269E-13 \\ \hline
\multirow{2}{*}{$^{238}$U} & 1.1445840000000025  & -44.79938793321383  & -44.760589846638325  & -2.6432998154494443  & -0.04828806750111746  & -0.19407065075590402 \\
 & 0.20311400000000246  & -1.344E-05  & -1.355E-05  & -1.553E-08  & -6.256E-12  & -1.314E-11 \\ \hline
\bottomrule
\end{longtable}

\begin{longtable}{
l
S[round-precision=3,table-format=1.3]
S[round-precision=2,table-format=5.2e1]
S[round-precision=2,table-format=5.2e1]
S[round-precision=2,table-format=5.2e1]
}
\caption{
\label{tab:2nu_2bminus_0_to_2}Values of $2\nu\bminusbminus$ PSFs in the closure and Taylor approximations for $0^{+}_{1}\to2^{+}_{1}$ transitions.}\\
\toprule
Nucleus & \ensuremath{Q\text{-value}}  & \ensuremath{G^{2\nu\beta^-\beta^-}\psfunit{-21}}  & \ensuremath{G_{22}\psfunit{-21}}  & \ensuremath{G_{6}\psfunit{-21}} \\ \midrule
\multirow{1}{*}{$^{46}$Ca} & 0.09937299999999971  & 4.454E-15  & 2.782E-16  & 6.210E-19 \\
\multirow{1}{*}{$^{48}$Ca} & 3.2845389999999997  & 4426.836846422774  & 265.5507123427349  & 601.0813508717184 \\
\multirow{1}{*}{$^{70}$Zn} & -0.04221800000001408  &    &    &   \\
\multirow{1}{*}{$^{76}$Ge} & 1.479956  & 0.46940604445107353  & 0.029145382886072938  & 0.014030286310143593 \\
\multirow{1}{*}{$^{80}$Se} & -0.482631999999992  &    &    &   \\
\multirow{1}{*}{$^{82}$Se} & 2.221374  & 84.3564105372918  & 5.203635030368105  & 5.57683257840586 \\
\multirow{1}{*}{$^{86}$Kr} & 0.180743840000003  & 1.787E-11  & 1.121E-12  & 8.205E-15 \\
\multirow{1}{*}{$^{94}$Zr} & 0.27364900000000314  & 2.727E-09  & 1.706E-10  & 2.857E-12 \\
\multirow{1}{*}{$^{96}$Zr} & 2.577791999999995  & 883.8284624975264  & 54.569522059034256  & 78.25426531803986 \\
\multirow{1}{*}{$^{98}$Mo} & -0.5435689999999967  &    &    &   \\
\multirow{1}{*}{$^{100}$Mo} & 2.4948897000000003  & 690.0318553115978  & 42.49166324343465  & 57.446027263220934 \\
\multirow{1}{*}{$^{104}$Ru} & 0.7435500000000006  & 3.648E-04  & 2.272E-05  & 2.809E-06 \\
\multirow{1}{*}{$^{110}$Pd} & 1.3600877000000002  & 0.5541341821946794  & 0.034598084546881604  & 0.014174714419041577 \\
\multirow{1}{*}{$^{114}$Cd} & -0.7551059999999932  &    &    &   \\
\multirow{1}{*}{$^{116}$Cd} & 1.5199399999999998  & 2.5456830652351488  & 0.15823337094314893  & 0.08096183774614613 \\
\multirow{1}{*}{$^{122}$Sn} & -0.19076499999998764  &    &    &   \\
\multirow{1}{*}{$^{124}$Sn} & 1.689939900000001  & 10.945356593099033  & 0.6815625754791351  & 0.4306870775391989 \\
\multirow{1}{*}{$^{128}$Te} & 0.4238292700000093  & 1.198E-06  & 7.457E-08  & 3.002E-09 \\
\multirow{1}{*}{$^{130}$Te} & 1.991442  & 97.0866027124493  & 6.047512275248301  & 5.266335463641782 \\
\multirow{1}{*}{$^{134}$Xe} & 0.21944526999999048  & 8.889E-10  & 5.533E-11  & 5.994E-13 \\
\multirow{1}{*}{$^{136}$Xe} & 1.639608  & 10.835654473649326  & 0.6760197750452287  & 0.40027566798885944 \\
\multirow{1}{*}{$^{142}$Ce} & -0.15862100000000035  &    &    &   \\
\multirow{1}{*}{$^{146}$Nd} & -0.6767299999999968  &    &    &   \\
\multirow{1}{*}{$^{148}$Nd} & 1.377745999999989  & 2.331571345902061  & 0.14484098514434865  & 0.061014613901134586 \\
\multirow{1}{*}{$^{150}$Nd} & 3.0374299999999947  & 37009.98140937972  & 2288.5683586292844  & 4553.795099310625 \\
\multirow{1}{*}{$^{154}$Sm} & 1.1277230999999943  & 0.26044960110377985  & 0.016259898157365203  & 0.004576599426414464 \\
\multirow{1}{*}{$^{160}$Gd} & 1.6435773000000053  & 28.20834899206847  & 1.7596469380064348  & 1.048081935993857 \\
\multirow{1}{*}{$^{170}$Er} & 0.5721603199999936  & 1.786E-04  & 1.121E-05  & 8.171E-07 \\
\multirow{1}{*}{$^{176}$Yb} & 0.9967569999999997  & 0.13481962463899655  & 0.008382676875577576  & 0.001855318287425209 \\
\multirow{1}{*}{$^{186}$W} & 0.3542800000000003  & 1.481E-06  & 9.231E-08  & 2.583E-09 \\
\multirow{1}{*}{$^{192}$Os} & 0.08971555000000159  & 4.489E-13  & 2.809E-14  & 5.058E-17 \\
\multirow{1}{*}{$^{198}$Pt} & 0.6384944899999987  & 0.0018089172987173485  & 1.133E-04  & 1.027E-05 \\
\multirow{1}{*}{$^{204}$Hg} & -0.47949800000000237  &    &    &   \\
\multirow{1}{*}{$^{232}$Th} & 0.7896919999999994  & 0.08043373109320617  & 0.005015527947901695  & 6.992E-04 \\
\multirow{1}{*}{$^{238}$U} & 1.1005190000000025  & 4.670480963124344  & 0.29189095467833  & 0.07882702313505276 \\
\midrule
Nucleus & \ensuremath{$Q$\text{-value}}  & \ensuremath{H^{2\nu\beta^-\beta^-}\psfunit{-21}}  & \ensuremath{H_{22}\psfunit{-21}}  & \ensuremath{H_{6}\psfunit{-21}} \\ \midrule
\multirow{1}{*}{$^{46}$Ca} & 0.09937299999999971  & -3.258E-17  & -6.114E-18  & -1.165E-20 \\
\multirow{1}{*}{$^{48}$Ca} & 3.2845389999999997  & -849.402673754375  & -153.5234015726666  & -322.0184947191016 \\
\multirow{1}{*}{$^{70}$Zn} & -0.04221800000001408  &    &    &   \\
\multirow{1}{*}{$^{76}$Ge} & 1.479956  & -0.05280426341280783  & -0.00984311378195067  & -0.004292661981601333 \\
\multirow{1}{*}{$^{80}$Se} & -0.482631999999992  &    &    &   \\
\multirow{1}{*}{$^{82}$Se} & 2.221374  & -12.408117659353893  & -2.301020735070182  & -2.2472719500638845 \\
\multirow{1}{*}{$^{86}$Kr} & 0.180743840000003  & -2.080E-13  & -3.931E-14  & -2.380E-16 \\
\multirow{1}{*}{$^{94}$Zr} & 0.27364900000000314  & -5.364E-11  & -1.006E-11  & -1.440E-13 \\
\multirow{1}{*}{$^{96}$Zr} & 2.577791999999995  & -140.173128003067  & -25.97226287013289  & -34.10857239299743 \\
\multirow{1}{*}{$^{98}$Mo} & -0.5435689999999967  &    &    &   \\
\multirow{1}{*}{$^{100}$Mo} & 2.4948897000000003  & -106.62552369969204  & -19.786634256363033  & -24.36998982119322 \\
\multirow{1}{*}{$^{104}$Ru} & 0.7435500000000006  & -2.155E-05  & -4.031E-06  & -4.419E-07 \\
\multirow{1}{*}{$^{110}$Pd} & 1.3600877000000002  & -0.05585007346904115  & -0.0104525460872422  & -0.00384121398441872 \\
\multirow{1}{*}{$^{114}$Cd} & -0.7551059999999932  &    &    &   \\
\multirow{1}{*}{$^{116}$Cd} & 1.5199399999999998  & -0.2757607369577115  & -0.051415636597706815  & -0.023690546592533158 \\
\multirow{1}{*}{$^{122}$Sn} & -0.19076499999998764  &    &    &   \\
\multirow{1}{*}{$^{124}$Sn} & 1.689939900000001  & -1.2777224499060358  & -0.23870475826006826  & -0.13525748133973214 \\
\multirow{1}{*}{$^{128}$Te} & 0.4238292700000093  & -3.589E-08  & -6.723E-09  & -2.318E-10 \\
\multirow{1}{*}{$^{130}$Te} & 1.991442  & -12.659115787674526  & -2.3631728363259037  & -1.8549569796969145 \\
\multirow{1}{*}{$^{134}$Xe} & 0.21944526999999048  & -9.868E-12  & -1.838E-12  & -1.517E-14 \\
\multirow{1}{*}{$^{136}$Xe} & 1.639608  & -1.217435789233687  & -0.22773843428937113  & -0.12109811594049971 \\
\multirow{1}{*}{$^{142}$Ce} & -0.15862100000000035  &    &    &   \\
\multirow{1}{*}{$^{146}$Nd} & -0.6767299999999968  &    &    &   \\
\multirow{1}{*}{$^{148}$Nd} & 1.377745999999989  & -0.22397656886581455  & -0.041919538566037494  & -0.01575229556696897 \\
\multirow{1}{*}{$^{150}$Nd} & 3.0374299999999947  & -6185.765174534458  & -1148.2236469935697  & -2096.2673683609396 \\
\multirow{1}{*}{$^{154}$Sm} & 1.1277230999999943  & -0.0208877904042013  & -0.003920156246591436  & -9.792E-04 \\
\multirow{1}{*}{$^{160}$Gd} & 1.6435773000000053  & -3.0891697680311516  & -0.5773635177065887  & -0.3083735930268848 \\
\multirow{1}{*}{$^{170}$Er} & 0.5721603199999936  & -6.811E-06  & -1.272E-06  & -7.860E-08 \\
\multirow{1}{*}{$^{176}$Yb} & 0.9967569999999997  & -0.009290497207328795  & -0.0017370281630244592  & -3.356E-04 \\
\multirow{1}{*}{$^{186}$W} & 0.3542800000000003  & -2.597E-08  & -4.862E-09  & -1.022E-10 \\
\multirow{1}{*}{$^{192}$Os} & 0.08971555000000159  & 1.776E-15  & 3.290E-16  & 8.019E-19 \\
\multirow{1}{*}{$^{198}$Pt} & 0.6384944899999987  & -7.235E-05  & -1.359E-05  & -1.028E-06 \\
\multirow{1}{*}{$^{204}$Hg} & -0.47949800000000237  &    &    &   \\
\multirow{1}{*}{$^{232}$Th} & 0.7896919999999994  & -0.0037764562493107882  & -7.129E-04  & -8.288E-05 \\
\multirow{1}{*}{$^{238}$U} & 1.1005190000000025  & -0.3180083880671525  & -0.05965551878590277  & -0.013742909340247902 \\
\bottomrule
\end{longtable}

\begin{longtable}{
l
S[round-precision=3,table-format=1.3]
S[round-precision=2,table-format=5.2e1]
S[round-precision=2,table-format=5.2e1]
}
\caption{\label{tab:0nu_2bminus_0_to_0}Values of $0\nu\beta^{-}\beta^{-}$ PSFs in the closure approximations for $0^{+}_{1}\to0^{+}_{1}$ and $0^{+}_{1}\to0^{+}_{2}$ transitions.}\\
\toprule
Nucleus & \ensuremath{Q\text{-value}}  & \ensuremath{G^{0\nu\bminusbminus}\psfunit{-15}}  & \ensuremath{H^{0\nu\bminusbminus}\psfunit{-15}} \\ \midrule
\multirow{2}{*}{$^{46}$Ca} & 0.9886589999999996  & 0.1473845492423125  & -0.09468741001646769 \\
 & -1.6223410000000005  &    &   \\ \hline
\multirow{2}{*}{$^{48}$Ca} & 4.26807  & 25.083645524476115  & -23.28276814379212 \\
 & 1.2707599999999997  & 0.3083380704695943  & -0.21868661010529047 \\ \hline
\multirow{2}{*}{$^{70}$Zn} & 0.997287999999986  & 0.23338045040973665  & -0.14874162934561902 \\
 & -0.2183330000000141  &    &   \\ \hline
\multirow{2}{*}{$^{76}$Ge} & 2.039059  & 2.4160841925180434  & -1.9636309081659953 \\
 & 0.9167799999999999  & 0.20231199794019042  & -0.12389002134540616 \\ \hline
\multirow{2}{*}{$^{80}$Se} & 0.13396800000000803  & 0.004648141688330233  & -7.078E-04 \\
 & -1.186541999999992  &    &   \\ \hline
\multirow{2}{*}{$^{82}$Se} & 2.9979  & 10.388105183393371  & -9.1494982940587 \\
 & 1.5102  & 0.9789918778049498  & -0.728521439865992 \\ \hline
\multirow{2}{*}{$^{86}$Kr} & 1.257423840000003  & 0.624435621187416  & -0.4349390327921369 \\
 & -0.8485761599999968  &    &   \\ \hline
\multirow{2}{*}{$^{94}$Zr} & 1.144747000000003  & 0.6056088168329663  & -0.4053742318444197 \\
 & -0.5969029999999971  &    &   \\ \hline
\multirow{2}{*}{$^{96}$Zr} & 3.356028999999995  & 21.10930261860547  & -18.908707657740923 \\
 & 2.207898999999995  & 4.730508585723573  & -3.905759704613007 \\ \hline
\multirow{2}{*}{$^{98}$Mo} & 0.10889100000000326  & 0.006542883406934676  & -7.798E-04 \\
 & -1.2132689999999968  &    &   \\ \hline
\multirow{2}{*}{$^{100}$Mo} & 3.0344  & 16.25086870625952  & -14.310618375661738 \\
 & 1.904095  & 3.239582535066339  & -2.573059665235489 \\ \hline
\multirow{2}{*}{$^{104}$Ru} & 1.2993600000000005  & 1.115887773788561  & -0.782702666424358 \\
 & -0.034229999999999317  &    &   \\ \hline
\multirow{2}{*}{$^{110}$Pd} & 2.01785  & 4.936211116023998  & -3.976787975269608 \\
 & 0.5447800000000003  & 0.135086240339498  & -0.06151850926026277 \\ \hline
\multirow{2}{*}{$^{114}$Cd} & 0.5448010000000068  & 0.15716041412738  & -0.07136294998803346 \\
 & -1.4084649999999934  &    &   \\ \hline
\multirow{2}{*}{$^{116}$Cd} & 2.8135  & 17.05788551480731  & -14.79597742377981 \\
 & 1.056636  & 0.7922664826906332  & -0.5099677761635745 \\ \hline
\multirow{2}{*}{$^{122}$Sn} & 0.37332900000001246  & 0.08257484909013403  & -0.028958040270743314 \\
 & -0.9840719999999876  &    &   \\ \hline
\multirow{2}{*}{$^{124}$Sn} & 2.292667000000001  & 9.344005849732078  & -7.755481054141385 \\
 & 0.6353840000000013  & 0.24903381228469593  & -0.12376289224690126 \\ \hline
\multirow{2}{*}{$^{128}$Te} & 0.8667402700000093  & 0.6143257728428537  & -0.35986390754417613 \\
 & -0.7162357299999909  &    &   \\ \hline
\multirow{2}{*}{$^{130}$Te} & 2.52751  & 14.58142739955254  & -12.363816865515892 \\
 & 0.7339899999999999  & 0.40302903322024486  & -0.21666579903990624 \\ \hline
\multirow{2}{*}{$^{134}$Xe} & 0.8241675699999905  & 0.6237626029108166  & -0.35576954781739584 \\
 & -0.9363874300000096  &    &   \\ \hline
\multirow{2}{*}{$^{136}$Xe} & 2.45813  & 14.95581659166488  & -12.596660156507914 \\
 & 0.8791610000000001  & 0.7247004547717589  & -0.4268481632097326 \\ \hline
\multirow{2}{*}{$^{142}$Ce} & 1.4171589999999996  & 3.616596731514861  & -2.598258335639922 \\
 & -0.8003250000000002  &    &   \\ \hline
\multirow{2}{*}{$^{146}$Nd} & 0.07044400000000314  & 0.017327262452050227  & -0.0010884577438869914 \\
 & -2.1405559999999966  &    &   \\ \hline
\multirow{2}{*}{$^{148}$Nd} & 1.9280009999999892  & 10.392738330072289  & -8.223419508156695 \\
 & 0.5035409999999891  & 0.3311715722950895  & -0.1411143745037815 \\ \hline
\multirow{2}{*}{$^{150}$Nd} & 3.3713849999999947  & 64.4917303175332  & -57.556174857845555 \\
 & 2.6309209999999945  & 27.835066693060593  & -23.74448711511657 \\ \hline
\multirow{2}{*}{$^{154}$Sm} & 1.2507939999999944  & 3.390432047584058  & -2.320316830710388 \\
 & 0.5701266999999944  & 0.5040198862975482  & -0.23204610252551758 \\ \hline
\multirow{2}{*}{$^{160}$Gd} & 1.7303650000000053  & 9.888296746537828  & -7.570610196850113 \\
 & 0.45042300000000535  & 0.368280312500298  & -0.14479840972650548 \\ \hline
\multirow{2}{*}{$^{170}$Er} & 0.6564149999999936  & 1.1329215118678206  & -0.5645690536176939 \\
 & -0.41293500000000616  &    &   \\ \hline
\multirow{2}{*}{$^{176}$Yb} & 1.0851059999999997  & 4.377922591677072  & -2.814884335207794 \\
 & -0.06483400000000028  &    &   \\ \hline
\multirow{2}{*}{$^{186}$W} & 0.4914300000000003  & 1.0681819591154584  & -0.4408004013279646 \\
 & -0.5695699999999997  &    &   \\ \hline
\multirow{2}{*}{$^{192}$Os} & 0.4062220000000016  & 0.9013370997619778  & -0.32386315444134284 \\
 & -0.7889469999999985  &    &   \\ \hline
\multirow{2}{*}{$^{198}$Pt} & 1.0502969999999987  & 7.997133050643778  & -5.046307751560906 \\
 & -0.3512230000000012  &    &   \\ \hline
\multirow{2}{*}{$^{204}$Hg} & 0.4196669999999976  & 1.4007822584945453  & -0.5128645729914083 \\
 & -1.1630330000000024  &    &   \\ \hline
\multirow{2}{*}{$^{232}$Th} & 0.8372649999999994  & 14.644344620533426  & -8.23495295204483 \\
 & 0.14584499999999945  & 0.8153585531338495  & -0.10585372152624334 \\ \hline
\multirow{2}{*}{$^{238}$U} & 1.1445840000000025  & 35.50998517915976  & -23.14534001291991 \\
 & 0.20311400000000246  & 1.5774079871144913  & -0.29567930984208807 \\ \hline
\bottomrule
\end{longtable}

\begin{longtable}{
l
S[round-precision=3,table-format=1.3]
S[round-precision=2,table-format=5.2e1]
S[round-precision=2,table-format=5.2e1]
}
\caption{\label{tab:2nu_2bplus_0_to_0}Values of $2\nu\beta^{+}\beta^{+}$ PSFs in the closure approximations for $0^{+}_{1}\to0^{+}_{1}$ and $0^{+}_{1}\to0^{+}_{2}$ transitions.}\\
\toprule
Nucleus & \ensuremath{Q\text{-value}}  & \ensuremath{G\twonutwobplus\psfunit{-29}}  & \ensuremath{H\twonutwobplus\psfunit{-29}} \\ \midrule
\multirow{2}{*}{$^{78}$Kr} & 0.8036732000000093  & 10390.408967761414  & -4243.851436223902 \\
 & 0.18994620000000928  & 1.011E-07  & -4.945E-09 \\ 
\multirow{2}{*}{$^{96}$Ru} & 0.6705082000000009  & 1098.971113470719  & -418.45944503736547 \\
 & -0.10772879999999907  &    &   \\ 
\multirow{2}{*}{$^{106}$Cd} & 0.7313941999999996  & 2116.0056887295764  & -859.6213506307822 \\
 & 0.21954419999999952  & 2.042E-07  & -1.207E-08 \\ 
\multirow{2}{*}{$^{124}$Xe} & 0.8127412000000085  & 4624.026506627647  & -2029.1458011297764 \\
 & 0.21001410000000853  & 5.367E-08  & -3.199E-09 \\ 
\multirow{2}{*}{$^{130}$Ba} & 0.5797022000000043  & 127.43143025593963  & -46.78598658656409 \\
 & 0.043634200000004286  & 1.815E-19  & -3.943E-21 \\ 
\multirow{2}{*}{$^{136}$Ce} & 0.3345331999999952  & 0.29905563482574177  & -0.07963052753642066 \\
 & -0.4839888000000049  &    &   \\ 
\bottomrule
\end{longtable}

\begin{longtable}{
l
S[round-precision=3,table-format=1.3]
S[round-precision=2,table-format=5.2e1]
S[round-precision=2,table-format=5.2e1]
}
\caption{\label{tab:0nu_2bplus_0_to_0}Values of $0\nu\beta^{+}\beta^{+}$ PSFs in the closure approximations for $0^{+}_{1}\to0^{+}_{1}$ transitions.}\\
\toprule
Nucleus & \ensuremath{Q\text{-value}}  & \ensuremath{G\twonutwobplus\psfunit{-20}}  & \ensuremath{H\twonutwobplus\psfunit{-20}} \\ \midrule
\multirow{1}{*}{$^{78}$Kr} & 0.8036732000000093  & 258.8185537275896  & -164.63933025174293 \\
\multirow{1}{*}{$^{96}$Ru} & 0.6705082000000009  & 87.43033454275019  & -51.84542155748696 \\
\multirow{1}{*}{$^{106}$Cd} & 0.7313941999999996  & 99.52750184856124  & -61.52156659868411 \\
\multirow{1}{*}{$^{124}$Xe} & 0.8127412000000085  & 113.20701414796041  & -73.4003535966261 \\
\multirow{1}{*}{$^{130}$Ba} & 0.5797022000000043  & 27.745888202605226  & -15.554053118302424 \\
\multirow{1}{*}{$^{136}$Ce} & 0.3345331999999952  & 2.5777901172043447  & -1.0779543583749696 \\
\bottomrule
\end{longtable}

\begin{longtable}{
l
S[round-precision=3,table-format=1.3]
S[round-precision=2,table-format=5.2e1]
}
\caption{\label{tab:2nu_ECbplus_0_to_0}Values of $2\nu\text{EC}\beta^{+}$ PSFs in the closure approximations for $0^{+}_{1}\to0^{+}_{1}$ and $0^{+}_{1}\to0^{+}_{2}$ transitions.}\\
\toprule
Nucleus & \ensuremath{Q\text{-value}}  & \ensuremath{G\twonuecbplus\psfunit{-24}} \\ \midrule
\multirow{2}{*}{$^{50}$Cr} & 0.14850509999999706  & 1.205E-06 \\
 & -3.719794900000003  &   \\ \hline
\multirow{2}{*}{$^{58}$Ni} & 0.9044021000000015  & 1.0424740978693947 \\
 & -1.3535478999999981  &   \\ \hline
\multirow{2}{*}{$^{64}$Zn} & 0.07301810000000342  & 3.765E-09 \\
 & -2.7943818999999968  &   \\ \hline
\multirow{2}{*}{$^{74}$Se} & 0.18724309999999456  & 1.048E-05 \\
 & -1.2955669000000054  &   \\ \hline
\multirow{2}{*}{$^{78}$Kr} & 1.8256711000000092  & 365.81782579931803 \\
 & 0.3270721000000092  & 9.355E-04 \\ \hline
\multirow{2}{*}{$^{84}$Sr} & 0.7677673700000036  & 0.6945033703922826 \\
 & -1.0695326299999963  &   \\ \hline
\multirow{2}{*}{$^{92}$Mo} & 0.6284391000000054  & 0.19687128956627387 \\
 & -0.7543308999999947  &   \\ \hline
\multirow{2}{*}{$^{96}$Ru} & 1.6925061000000008  & 390.7580628329422 \\
 & 0.5443761000000007  & 0.0726919919081284 \\ \hline
\multirow{2}{*}{$^{102}$Pd} & 0.18147409999999442  & 8.881E-06 \\
 & -0.7622159000000056  &   \\ \hline
\multirow{2}{*}{$^{106}$Cd} & 1.7533920999999995  & 678.6450073270726 \\
 & 0.6196320999999994  & 0.2450016566014011 \\ \hline
\multirow{2}{*}{$^{120}$Te} & 0.713583099999991  & 0.9084681421302643 \\
 & -1.161524900000009  &   \\ \hline
\multirow{2}{*}{$^{124}$Xe} & 1.8347391000000084  & 1441.9578601704343 \\
 & 0.17745610000000855  & 6.732E-06 \\ \hline
\multirow{2}{*}{$^{130}$Ba} & 1.6017001000000042  & 587.4627389350434 \\
 & -0.19181989999999582  &   \\ \hline
\multirow{2}{*}{$^{136}$Ce} & 1.356531099999995  & 187.43811554819348 \\
 & -0.22243790000000496  &   \\ \hline
\multirow{2}{*}{$^{144}$Sm} & 0.7604041000000019  & 2.5052986567491566 \\
 & -1.3242758999999977  &   \\ \hline
\multirow{2}{*}{$^{156}$Dy} & 0.9839491  & 24.41167120579065 \\
 & -0.06553790000000004  &   \\ \hline
\multirow{2}{*}{$^{162}$Er} & 0.8249610999999881  & 6.55218562485735 \\
 & -0.5752989000000119  &   \\ \hline
\multirow{2}{*}{$^{168}$Yb} & 0.3873661000000015  & 0.010711309439342056 \\
 & -0.8298028999999987  &   \\ \hline
\multirow{2}{*}{$^{174}$Hf} & 0.07794010000000196  & 2.167E-09 \\
 & -1.409179899999998  &   \\ \hline
\multirow{2}{*}{$^{184}$Os} & 0.43089009999999917  & 0.03336585244706995 \\
 & -0.5715999000000009  &   \\ \hline
\multirow{2}{*}{$^{190}$Pt} & 0.37932309999999636  & 0.010765501251474694 \\
 & -0.5324769000000036  &   \\ \hline
\bottomrule
\end{longtable}

\begin{longtable}{
l
S[round-precision=3,table-format=1.3]
S[round-precision=2,table-format=5.2e1]
}
\caption{\label{tab:0nu_ECbplus_0_to_0}Values of $0\nu\text{EC}\beta^{+}$ PSFs in the closure approximations for $0^{+}_{1}\to0^{+}_{1}$ and $0^{+}_{1}\to0^{+}_{2}$ transitions.}\\
\toprule
Nucleus & \ensuremath{Q\text{-value}}  & \ensuremath{G\zeronuecbplus\psfunit{-18}} \\ \midrule
\multirow{2}{*}{$^{50}$Cr} & 0.14850509999999706  & 0.02282648938219251 \\
 & -3.719794900000003  &   \\ \hline
\multirow{2}{*}{$^{58}$Ni} & 0.9044021000000015  & 0.7931739600605635 \\
 & -1.3535478999999981  &   \\ \hline
\multirow{2}{*}{$^{64}$Zn} & 0.07301810000000342  & 0.00970427899816383 \\
 & -2.7943818999999968  &   \\ \hline
\multirow{2}{*}{$^{74}$Se} & 0.18724309999999456  & 0.07517852990966052 \\
 & -1.2955669000000054  &   \\ \hline
\multirow{2}{*}{$^{78}$Kr} & 1.8256711000000092  & 5.108316074210908 \\
 & 0.3270721000000092  & 0.23997590195095322 \\ \hline
\multirow{2}{*}{$^{84}$Sr} & 0.7677673700000036  & 1.2431992190396817 \\
 & -1.0695326299999963  &   \\ \hline
\multirow{2}{*}{$^{92}$Mo} & 0.6284391000000054  & 1.1662129360807973 \\
 & -0.7543308999999947  &   \\ \hline
\multirow{2}{*}{$^{96}$Ru} & 1.6925061000000008  & 7.751690546068307 \\
 & 0.5443761000000007  & 1.0413416299992169 \\ \hline
\multirow{2}{*}{$^{102}$Pd} & 0.18147409999999442  & 0.1603074283827813 \\
 & -0.7622159000000056  &   \\ \hline
\multirow{2}{*}{$^{106}$Cd} & 1.7533920999999995  & 10.65565166850074 \\
 & 0.6196320999999994  & 1.7014505258366144 \\ \hline
\multirow{2}{*}{$^{120}$Te} & 0.713583099999991  & 2.755894687283658 \\
 & -1.161524900000009  &   \\ \hline
\multirow{2}{*}{$^{124}$Xe} & 1.8347391000000084  & 16.49439128779941 \\
 & 0.17745610000000855  & 0.25660051682112966 \\ \hline
\multirow{2}{*}{$^{130}$Ba} & 1.6017001000000042  & 14.631650441801833 \\
 & -0.19181989999999582  &   \\ \hline
\multirow{2}{*}{$^{136}$Ce} & 1.356531099999995  & 12.354809228636725 \\
 & -0.22243790000000496  &   \\ \hline
\multirow{2}{*}{$^{144}$Sm} & 0.7604041000000019  & 5.941443508196129 \\
 & -1.3242758999999977  &   \\ \hline
\multirow{2}{*}{$^{156}$Dy} & 0.9839491  & 11.897514639864513 \\
 & -0.06553790000000004  &   \\ \hline
\multirow{2}{*}{$^{162}$Er} & 0.8249610999999881  & 10.014492839034027 \\
 & -0.5752989000000119  &   \\ \hline
\multirow{2}{*}{$^{168}$Yb} & 0.3873661000000015  & 3.1047449393750908 \\
 & -0.8298028999999987  &   \\ \hline
\multirow{2}{*}{$^{174}$Hf} & 0.07794010000000196  & 0.21457115362919002 \\
 & -1.409179899999998  &   \\ \hline
\multirow{2}{*}{$^{184}$Os} & 0.43089009999999917  & 5.727887781603741 \\
 & -0.5715999000000009  &   \\ \hline
\multirow{2}{*}{$^{190}$Pt} & 0.37932309999999636  & 5.317456030932652 \\
 & -0.5324769000000036  &   \\ \hline
\bottomrule
\end{longtable}

\begin{longtable}{
l
S[round-precision=3,table-format=1.3]
S[round-precision=2,table-format=5.2e1]
S[round-precision=2,table-format=5.2e1]
S[round-precision=2,table-format=5.2e1]
S[round-precision=2,table-format=5.2e1]
S[round-precision=2,table-format=5.2e1]
}
\caption{\label{tab:2nu_2EC_0_to_0}Values of $2\nu\text{ECEC}$ PSFs in the closure and Taylor approximations for $0^{+}_{1}\to0^{+}_{1}$ and $0^{+}_{1}\to0^{+}_{2}$ transitions.}\\
\toprule
Nucleus & \ensuremath{Q\text{-value}}  & \ensuremath{G\twonutwoec \psfunit{-24}}  & \ensuremath{G_{0}\psfunit{-24}}  & \ensuremath{G_{2}\psfunit{-24}}  & \ensuremath{G_{22}\psfunit{-24}}  & \ensuremath{G_{4}\psfunit{-24}} \\ \midrule
\multirow{2}{*}{$^{36}$Ar} & 0.43259899999999835  & 4.274E-04  & 4.273E-04  & 5.322E-06  & 3.867E-08  & 7.735E-08 \\
 & -2.9134010000000017  &    &    &    &    &   \\ \hline
\multirow{2}{*}{$^{40}$Ca} & 0.19349796999999672  & 1.358E-05  & 1.358E-05  & 3.223E-08  & 4.456E-11  & 8.928E-11 \\
 & -1.927412030000003  &    &    &    &    &   \\ \hline
\multirow{2}{*}{$^{50}$Cr} & 1.170502999999997  & 0.43380310778696224  & 0.4331378812307765  & 0.03983907459302071  & 0.0021373904784051964  & 0.004275168452641435 \\
 & -2.697797000000003  &    &    &    &    &   \\ \hline
\multirow{2}{*}{$^{54}$Fe} & 0.6807640000000028  & 0.04760496020397274  & 0.047582549381627336  & 0.0014522108636139746  & 2.585E-05  & 5.172E-05 \\
 & -2.1488559999999968  &    &    &    &    &   \\ \hline
\multirow{2}{*}{$^{58}$Ni} & 1.9264000000000014  & 15.355471807607977  & 15.30024230005167  & 3.8229330735417606  & 0.55717708970351  & 1.1144287221191425 \\
 & -0.33154999999999824  &    &    &    &    &   \\ \hline
\multirow{2}{*}{$^{64}$Zn} & 1.0950160000000033  & 1.4202943563325425  & 1.4188252844223184  & 0.11271260535697528  & 0.005222178944193731  & 0.010447418478223634 \\
 & -1.7723839999999966  &    &    &    &    &   \\ \hline
\multirow{2}{*}{$^{74}$Se} & 1.2092409999999945  & 5.705896004170501  & 5.699706708764409  & 0.5490147746631788  & 0.030839907606998472  & 0.06170645495229278 \\
 & -0.2735690000000055  &    &    &    &    &   \\ \hline
\multirow{2}{*}{$^{78}$Kr} & 2.847669000000009  & 668.6681045721984  & 664.762920433538  & 362.0932748275855  & 115.04402380992975  & 230.11028283198505 \\
 & 1.349070000000009  & 15.102930605548256  & 15.083591755748419  & 1.8075377421424064  & 0.12631711080304378  & 0.25274821656877816 \\ \hline
\multirow{2}{*}{$^{84}$Sr} & 1.7897652700000035  & 95.18214522920248  & 94.98166547727668  & 20.15037961635864  & 2.493179294547136  & 4.987979928690853 \\
 & -0.04753472999999642  &    &    &    &    &   \\ \hline
\multirow{2}{*}{$^{92}$Mo} & 1.6504370000000053  & 135.32518674980233  & 135.10636553961655  & 24.109219574822156  & 2.508625973986557  & 5.020371141390252 \\
 & 0.2676670000000052  & 0.008416988775900386  & 0.0084167001058981  & 3.189E-05  & 6.886E-08  & 1.432E-07 \\ \hline
\multirow{2}{*}{$^{96}$Ru} & 2.7145040000000007  & 2465.24436723838  & 2454.7424406905866  & 1201.7980517797423  & 343.1607115543101  & 686.510521009495 \\
 & 1.5663740000000006  & 149.04654179885927  & 148.83998792023945  & 23.756716257097118  & 2.210706046356301  & 4.425254611340857 \\ \hline
\multirow{2}{*}{$^{102}$Pd} & 1.2034719999999943  & 54.37281010736254  & 54.33185212993968  & 5.010738189126149  & 0.26924559454584013  & 0.5395112765783231 \\
 & 0.2597819999999943  & 0.012654743076158995  & 0.012654394344450906  & 4.272E-05  & 8.060E-08  & 1.733E-07 \\ \hline
\multirow{2}{*}{$^{106}$Cd} & 2.7753899999999994  & 5608.183924282443  & 5585.684271427698  & 2844.213501282438  & 844.6068837634032  & 1689.887181154363 \\
 & 1.6416299999999993  & 381.5284157032451  & 381.0062965056618  & 66.30869181393676  & 6.726600924288864  & 13.46938296410317 \\ \hline
\multirow{2}{*}{$^{108}$Cd} & 0.27179200000000125  & 0.021746218315406365  & 0.021745605772160337  & 7.945E-05  & 1.612E-07  & 3.502E-07 \\
 & -0.7809879999999988  &    &    &    &    &   \\ \hline
\multirow{2}{*}{$^{120}$Te} & 1.735580999999991  & 989.2067238027703  & 987.8794016402071  & 190.84146362227423  & 21.484410866830945  & 43.03710907307847 \\
 & -0.13952700000000906  &    &    &    &    &   \\ \hline
\multirow{2}{*}{$^{124}$Xe} & 2.8567370000000083  & 17801.078974701282  & 17736.959893725078  & 9488.598735810037  & 2959.7719039184594  & 5923.486063087702 \\
 & 1.1994540000000085  & 199.23929339125797  & 199.1201785650303  & 17.720195784223563  & 0.9174245373645122  & 1.8427766228394729 \\ \hline
\multirow{2}{*}{$^{126}$Xe} & 0.917781130000003  & 48.12905699122411  & 48.11297565310111  & 2.4320597345063883  & 0.07136448230296373  & 0.14386025142350617 \\
 & -0.955609869999997  &    &    &    &    &   \\ \hline
\multirow{2}{*}{$^{130}$Ba} & 2.623698000000004  & 15919.876975941612  & 15874.108365566084  & 7109.817644861695  & 1856.4443284318863  & 3716.433032555614 \\
 & 0.8301780000000041  & 38.06164093545651  & 38.05180216911833  & 1.5354048034211736  & 0.03587145147498847  & 0.0726185676578405 \\ \hline
\multirow{2}{*}{$^{132}$Ba} & 0.8440715099999943  & 41.59007046216241  & 41.57909786870433  & 1.738673663587186  & 0.04210765909095702  & 0.0852037366983924 \\
 & 0.8440715099999943  & 41.59007046216241  & 41.57909786870433  & 1.738673663587186  & 0.04210765909095702  & 0.0852037366983924 \\ \hline
\multirow{2}{*}{$^{136}$Ce} & 2.378528999999995  & 13226.284483862568  & 13196.705224372121  & 4812.38140717842  & 1022.7921852353119  & 2048.4500721986296 \\
 & 0.7995599999999949  & 41.734204023037684  & 41.72480653291106  & 1.5343163173307455  & 0.032589762626102414  & 0.06623536906487619 \\ \hline
\multirow{2}{*}{$^{138}$Ce} & 0.6959389999999985  & 19.409227773491804  & 19.4060439725201  & 0.5275159132708219  & 0.008251227827314585  & 0.016877144551696095 \\
 & -1.6440610000000013  &    &    &    &    &   \\ \hline
\multirow{2}{*}{$^{144}$Sm} & 1.7824020000000018  & 5505.959089004924  & 5499.63615464483  & 1091.2972800335333  & 126.02951209016925  & 252.98540868828107 \\
 & -0.3022779999999978  &    &    &    &    &   \\ \hline
\multirow{2}{*}{$^{152}$Gd} & 0.055672999999995226  & 2.273E-06  & 2.273E-06  & 3.113E-10  & 2.231E-14  & 5.454E-14 \\
 & -0.6290780000000048  &    &    &    &    &   \\ \hline
\multirow{2}{*}{$^{156}$Dy} & 2.005947  & 18687.1872494705  & 18662.1273892829  & 4684.015779165316  & 684.1149888449007  & 1373.5981410127565 \\
 & 0.9564599999999999  & 352.07850929049073  & 351.9811679851752  & 18.228308989224985  & 0.5435097707121007  & 1.1105845448917844 \\ \hline
\multirow{2}{*}{$^{158}$Dy} & 0.28281699999999543  & 0.21600483303568827  & 0.2160007717291508  & 7.706E-04  & 1.189E-06  & 3.738E-06 \\
 & -0.9133470000000046  &    &    &    &    &   \\ \hline
\multirow{2}{*}{$^{162}$Er} & 1.846958999999988  & 16301.885037912356  & 16284.294107873826  & 3416.0018320992463  & 416.5768724848141  & 837.753186600671 \\
 & 0.44669899999998797  & 5.378180055253607  & 5.377921204711599  & 0.050357206092715684  & 2.472E-04  & 5.897E-04 \\ \hline
\multirow{2}{*}{$^{164}$Er} & 0.02507600000000093  & 1.144E-08  & 1.144E-08  & 4.112E-13  & 5.138E-18  & 1.692E-17 \\
 & -1.6296339999999991  &    &    &    &    &   \\ \hline
\multirow{2}{*}{$^{168}$Yb} & 1.4093640000000014  & 5157.074142910706  & 5154.075048473325  & 604.4938736271356  & 41.054563871363456  & 83.09318154471781 \\
 & 0.19219500000000123  & 0.019078948375993228  & 0.019078749175815892  & 4.019E-05  & 2.428E-08  & 1.034E-07 \\ \hline
\multirow{2}{*}{$^{174}$Hf} & 1.0999380000000019  & 1765.3255341492466  & 1764.7513266845317  & 119.92372821209678  & 4.681623034091564  & 9.601319708368846 \\
 & -0.38718199999999814  &    &    &    &    &   \\ \hline
\multirow{2}{*}{$^{180}$W} & 0.14323300000000017  & 0.0035012501221365153  & 0.003501229444393574  & 4.470E-06  & 2.230E-09  & 6.196E-09 \\
 & -0.9586669999999999  &    &    &    &    &   \\ \hline
\multirow{2}{*}{$^{184}$Os} & 1.452887999999999  & 14325.791361723812  & 14317.882037789332  & 1746.1093996215247  & 122.8810280917646  & 250.13231257704697 \\
 & 0.45039799999999897  & 14.07688912157355  & 14.076301320232574  & 0.12988226807466385  & 5.693E-04  & 0.0015710642180419721 \\ \hline
\multirow{2}{*}{$^{190}$Pt} & 1.4013209999999963  & 15579.419995220604  & 15571.788054022985  & 1740.0027421020393  & 111.88052603785226  & 228.80638964026986 \\
 & 0.4895209999999963  & 30.386389767230174  & 30.384930161693536  & 0.333031675959104  & 0.0017549658079098964  & 0.004757623362168179 \\ \hline
\multirow{2}{*}{$^{196}$Hg} & 0.8185940000000009  & 949.9447078761397  & 949.8087749457543  & 31.989313107260053  & 0.592314211245368  & 1.3064426646060057 \\
 & -0.31671799999999894  &    &    &    &    &   \\ \hline
\bottomrule
\end{longtable}

\twocolumn

\section{Conclusion}\label{sec13}

We present a comprehensive set of calculations for the kinematic factors involved in DBD, namely PSFs, electron spectra and angular correlations for the channels $\bminusbminus$, $\beta^+\beta^+$, EC$\beta^+$ and ECEC and transitions to final g.s. and first excited states $2^+$ and $0^+$.  The theoretical formulas are derived in closure approximation and with a Taylor expansion method. The calculations are performed with an adapted DHFS method and using the \verb|RADIAL| computing package.The key features of our approach are: the accurate treatment of electron and positron wave functions in bound and continuous states, inclusion of all relevant atomic ingredients like screening, exchange, radiative, finite nuclear size and phase shift corrections. Our study extends on a large number of nuclei which can undergo double-beta decays. Our extended study of the impact that each atomic correction has on the results confirms the previous reports. Our obtained PSF values are generally in the line with other results from literature, but for particular transitions and cases there are relevant differences which can be useful for accurate and extended analyses of the DBD data. We also provide numerical values for electron energy spectra and angular correlations for isotopes $\nucleus{76}{Ge}$, $\nucleus{130}{Te}$ and $\nucleus{136}{Xe}$, while similar data for other isotopes can also be provided upon request.  

\backmatter

\bmhead{Acknowledgements}
We acknowledge support from project PNRR-I8/C9-CF264, Contract No. 760100/23.05.2023 of the Romanian National Authority for Research. We also wish to thank O.V. Ni\c{t}escu and F. \v{S}imkovic for valuable and fruitful discussions.

\bibliography{bibliography}

\end{document}